\documentclass{emulateapj-rtx4}
\usepackage{lscape}

\submitted{Accepted for Publication in the Astronomical Journal: June 22,
2015}
\shorttitle{NGC 6273 Abundances}
\shortauthors{Johnson et al.}

\newcommand\iso[2]{$^{\rm #1}$#2}

\begin{document}

\title{A SPECTROSCOPIC ANALYSIS OF THE GALACTIC GLOBULAR CLUSTER
NGC 6273 (M19)\footnote{This paper includes data gathered with the 6.5 meter 
Magellan Telescopes located at Las Campanas Observatory, Chile.}}

\author{
Christian I. Johnson\altaffilmark{1,2},
R. Michael Rich\altaffilmark{3},
Catherine A. Pilachowski\altaffilmark{4},
Nelson Caldwell\altaffilmark{1},
Mario Mateo\altaffilmark{5}, 
John I. Bailey, III\altaffilmark{5}, and
Jeffrey D. Crane\altaffilmark{6}
}

\altaffiltext{1}{Harvard--Smithsonian Center for Astrophysics, 60 Garden
Street, MS--15, Cambridge, MA 02138, USA; cjohnson@cfa.harvard.edu; 
ncaldwell@cfa.harvard.edu}

\altaffiltext{2}{Clay Fellow}

\altaffiltext{3}{Department of Physics and Astronomy, UCLA, 430 Portola Plaza,
Box 951547, Los Angeles, CA 90095-1547, USA; rmr@astro.ucla.edu}

\altaffiltext{4}{Astronomy Department, Indiana University Bloomington, Swain
West 319, 727 East 3rd Street, Bloomington, IN 47405--7105, USA;
catyp@astro.indiana.edu}

\altaffiltext{5}{Department of Astronomy, University of Michigan, Ann Arbor, 
MI 48109, USA; mmateo@umich.edu; baileyji@umich.edu}

\altaffiltext{6}{The Observatories of the Carnegie Institution for Science,
Pasadena, CA 91101, USA; crane@obs.carnegiescience.edu}

\begin{abstract}

A combined effort utilizing spectroscopy and photometry has revealed
the existence of a new globular cluster class.  These ``anomalous" clusters,
which we refer to as ``iron--complex" clusters, are differentiated from normal 
clusters by exhibiting large ($\ga$0.10 dex) intrinsic metallicity dispersions,
complex sub--giant branches, and correlated [Fe/H] and s--process enhancements.
In order to further investigate this phenomenon, we have measured radial 
velocities and chemical abundances for red giant branch stars in the massive, 
but scarcely studied, globular cluster NGC 6273.  The velocities and abundances
were determined using high resolution (R$\sim$27,000) spectra obtained with the
Michigan/Magellan Fiber System (M2FS) and MSpec spectrograph on the 
Magellan--Clay 6.5m telescope at Las Campanas Observatory.  We find that NGC 
6273 has an average heliocentric radial velocity of $+$144.49 km s$^{\rm -1}$ 
($\sigma$=9.64 km s$^{\rm -1}$) and an extended metallicity distribution 
([Fe/H]=--1.80 to --1.30) composed of at least two distinct stellar 
populations.  Although the two dominant populations have similar [Na/Fe], 
[Al/Fe], and [$\alpha$/Fe] abundance patterns, the more metal--rich stars 
exhibit significant [La/Fe] enhancements.  The [La/Eu] data indicate that the 
increase in [La/Fe] is due to almost pure s--process enrichment.  A third more 
metal--rich population with low [X/Fe] ratios may also be present.  Therefore, 
NGC 6273 joins clusters such as $\omega$ Centauri, M 2, M 22, and NGC 5286 as 
a new class of iron--complex clusters exhibiting complicated star formation
histories.

\end{abstract}

\keywords{stars: abundances, globular clusters: general, globular clusters:
individual (NGC 6273, M 19)}

\section{INTRODUCTION}

Galactic globular clusters are proving to be a rich and diverse population.  
These objects are generally older than about 10 Gyr ($\sigma$$\sim$3 Gyr; 
e.g., Rosenberg et al. 1999; Salaris \& Weiss 2002; De Angeli et al. 2005; 
Mar{\'{\i}}n--Franch et al. 2009; VandenBergh et al. 2013), but also span about 
a factor of 300 in metallicity (e.g., Zinn \& West 1984; Harris 1996; Carretta 
\& Gratton 1997; Kraft \& Ivans 2003; Carretta et al. 2009a).  Early high 
resolution spectroscopic work revealed that clusters typically contain stars 
with similar heavy element abundances, most notably [Fe/H]\footnote{[A/B]$\equiv$log(N$_{\rm A}$/N$_{\rm B}$)$_{\rm star}$--log(N$_{\rm A}$/N$_{\rm B}$)$_{\sun}$ and log $\epsilon$(A)$\equiv$log(N$_{\rm A}$/N$_{\rm H}$)+12.0 for elements A and B.}, but with $>$0.5--1.0 dex star--to--star abundance variations for 
elements lighter than about Si (e.g., Cohen 1978; Peterson 1980; Sneden et al. 
1991; Kraft et al. 1992; Pilachowski et al. 1996; Gratton et al. 2001).  
Furthermore, it was discovered that the light element abundance variations are 
(anti--)correlated with one another, and that the correlation patterns, such as
the O--Na anti--correlation and Na--Al correlation, are evidence that the gas 
from which the present--day low mass globular cluster stars formed was 
subjected to high--temperature proton--capture nucleosynthesis (e.g., 
Denisenkov \& Denisenkova 1990; Langer et al. 1993; Prantzos et al. 2007).  
Except for a few notable cases (Koch et al. 2009; Caloi \& D'Antona 2011; 
Villanova et al. 2013), recent large sample spectroscopic surveys (e.g., 
Carretta et al. 2009b,c) have built upon this early work and cemented the idea 
that the correlated light element abundance variations are a characteristic 
common to perhaps all Galactic globular clusters (see also reviews by Kraft 
1994; Gratton et al. 2004; 2012).  These spectroscopic observations provided 
the first evidence that globular clusters may not be simple stellar populations.

Concurrent with spectroscopic work has been the revelation that, when observed
using appropriate filter combinations, the color--magnitude diagrams of many 
globular clusters exhibit multiple, often discreet, photometric sequences that
can extend from the main--sequence to the asymptotic giant branch (AGB; e.g., 
Piotto et al. 2007, 2012, 2015; Lardo et al. 2011; Milone et al. 2012a,b, 2015; 
Monelli et al. 2013; Cummings et al. 2014; Lim et al. 2015).  Since many of 
the filters used are sensitive to a star's light element composition (e.g.,
Bond \& Neff 1969), the combined spectroscopic evidence of large star--to--star
light element abundance variations and photometric evidence of multiple 
color--magnitude diagram sequences reveals that nearly all clusters host more 
than one stellar population.  For clusters exhibiting a negligible spread in 
[Fe/H], the various populations are often categorized by their light element 
chemistry as ``primordial", ``intermediate", or ``extreme" stars (e.g., 
Carretta et al. 2009c).  The primordial stars are thought to be the first 
generation and have a composition similar to halo field stars (high [O/Fe] and 
low [Na/Fe]), and the intermediate stars are thought to be second generation 
stars that exhibit lower [O/Fe] and higher [Na/Fe].  Only a small number of 
clusters host extreme stars, which are distinguished as having 
[Na/Fe]$\ga$$+$0.4 and [O/Fe]$\la$--0.2.  The intermediate population 
dominates by number in most clusters (Carretta et al. 2009c), and the 
intermediate and extreme stars may also have enhanced He relative to the 
primordial population (e.g., Bragaglia et al. 2010; Dupree et al. 2011; 
Pasquini et al. 2011; Mucciarelli et al. 2014).

While spectroscopic and photometric analyses provide clear evidence that 
more than one stellar population, differentiated by light element chemistry, 
exists within most globular clusters, no consensus has yet been reached 
regarding the cause of the light element variations nor its interpretation.
The sole stable argument in the debate about globular cluster composition
is that, outside of normal dredge--up processes, the light element abundance 
variations were largely already imprinted on the gas from which the stars 
formed.  This result is most clearly evidenced by observations of unevolved 
main--sequence and sub--giant branch (SGB) stars exhibiting the same light 
element correlations as the red giant branch (RGB) stars (e.g., Briley et al. 
1996; Gratton et al. 2001; Cohen \& Mel{\'e}ndez 2005; Bragaglia et al. 2010; 
D'Orazi et al. 2010; Dobrovolskas et al. 2014).  However, the available data
make differentiating various pollution scenarios difficult.  To date,
none of the proposed nucleosynthesis sources, which include intermediate mass
($\sim$5--8 M$_{\rm \odot}$) AGB stars (e.g., Fenner et al. 2004; Karakas et 
al. 2006; Ventura \& D'Antona 2009; D'Ercole et al. 2010; Ventura et al.
2013; Doherty et al. 2014), massive rapidly rotating main--sequence stars
(e.g., Decressin et al. 2007; 2010), interacting massive binary stars (de Mink
et al. 2009; Izzard et al. 2013), and very massive ($\sim$10$^{\rm 4}$ 
M$_{\rm \odot}$) stars (Denissenkov \& Hartwick 2014), are able to fully 
explain all observed abundance patterns.  Additionally, no combination of the 
previously proposed sources seems able to reproduce all abundance patterns 
either (Bastian et al. 2015).  A formal merging of the spectroscopic and 
photometric observations with theoretical models remains a work in progress.

Although the measured [Fe/H] dispersion is $\la$12$\%$ ($\la$0.05 dex) for 
many globular clusters (Carretta et al. 2009a), a population of about 8 known 
clusters exists for which the derived [Fe/H] spread exceeds the measurement 
errors (e.g., see Marino et al. 2015; their Table 10).  These ``anomalous" 
clusters, which we refer to as ``iron--complex" clusters\footnote{The term 
``iron--complex" refers to any globular cluster exhibiting a significant 
($\ga$0.10 dex) [Fe/H] dispersion when measured from high resolution spectra.  
We have adopted this term in order to avoid confusing the word ``anomalous", 
which can refer to either a cluster with a metallicity dispersion (e.g., Marino
et al. 2015) or a sub--population with peculiar chemistry residing in a cluster
(e.g., Pancino et al. 2000; Yong et al. 2014).  We note that clusters with both 
[Fe/H] and s--process abundance spreads have also been referred to as 
``s--Fe--anomalous" (Marino et al. 2015).  However, we have avoided 
discriminating between the two subsets here because some of the clusters 
identified as anomalous have not yet had their heavy element abundances 
measured, and may in fact also be s--Fe--anomalous.}, are often identified by 
the following features: (1) a dispersion in [Fe/H] exceeding $\sim$0.10 dex 
when measured using moderately high dispersion and signal--to--noise ratio 
(S/N) spectra, (2) multiple photometric sequences, especially on the SGB, and 
(3) a significant abundance spread for light elements and also heavy elements 
that, in the Solar System, are produced by the slow neutron--capture process 
(s--process).  Many iron--complex clusters are also relatively massive and tend
to host very blue horizontal branches (HB).  The massive globular cluster omega
Centauri ($\omega$ Cen) is the best known and most extreme object from this 
group, and has been demonstrated by multiple authors to possess: a very blue 
HB, a range in [Fe/H] that spans about a factor of 100, at least five distinct 
main stellar populations (each with its own set of primordial, intermediate, 
and extreme stars), and a strong correlation between metallicity and elements 
such as Ba and La that are likely produced by the s--process (e.g., Norris \& 
Da Costa 1995; Suntzeff \& Kraft 1996; Lee et al. 1999; Smith et al. 2000; 
Bellini et al. 2010; Johnson \& Pilachowski 2010; D'Orazi et al. 2011; Marino 
et al. 2011a; Pancino et al. 2011; Villanova et al. 2014).  Less extreme 
examples also include M 22, M 2, M 54, NGC 1851, NGC 5286, NGC 5824, and Terzan
5 (M 22: e.g,. Hesser et al. 1977; Pilachowski et al. 1982; Lehnert et al. 
1991; Marino et al. 2009, 2011b, 2013; Da Costa et al. 2009; Roederer et al. 
2011; Alves--Brito et al. 2012; M 2: Piotto et al. 2012; Lardo et al. 2013; 
Yong et al. 2014; Milone et al. 2015; M 54: e.g., Sarajedini \& Layden 1995; 
Brown et al. 1999; Siegel et al. 2007; Bellazzini et al. 2008; Carretta et al. 
2010a; NGC 1851: e.g., Yong \& Grundahl 2008; Milone et al. 2009; Yong et al. 
2009, 2015; Zoccali et al. 2009; Carretta et al. 2010b, 2011; NGC 5286: Marino 
et al. 2015; NGC 5824: Saviane et al. 2012; Da Costa et al. 2014; Terzan 5: 
Ferraro et al. 2009; Origlia et al. 2011, 2013; Massari et al. 
2014)\footnote{We note that Simmerer et al. (2013) also found evidence 
supporting a metallicity spread in NGC 3201.  However, this claim is not 
supported by the observations presented in Mu{\~n}oz et al. (2013) and 
Mucciarelli et al. (2015) so we have omitted NGC 3201 from the list of 
iron--complex clusters.}.  Among these clusters, $\omega$ Cen, M 22, M 2, and 
NGC 5286 stand out because each has been confirmed to host at least two stellar
populations distinguished by their [Fe/H] and s--process abundances.

In this context, we examine here the chemical composition of RGB stars in the
globular cluster NGC 6273 (M19), which is a scarcely studied cluster near the 
Galactic bulge.  NGC 6273 is one of the most massive and luminous clusters in 
the Galaxy (M$_{\rm V}$=--9.13; M$\sim$1.2--1.6$\times$10$^{\rm 6}$ 
M$_{\rm \odot}$; Harris 1996; Gnedin \& Ostriker 1997; Brown et al. 2010), and 
despite suffering from significant differential reddening has shown some 
evidence supporting the existence of a metallicity spread and complex 
color--magnitude diagram (Harris et al. 1976; Rutledge et al. 1997; Piotto et 
al. 1999, 2002; Brown et al. 2010; Alonso--Garc{\'{\i}}a et al. 2012).  A
possible spread in color on the RGB and an extended multimodal blue HB are 
two particularly noteworthy features.  NGC 6273 is metal--poor 
([Fe/H]$\sim$--1.75), may be the most elliptical cluster in the Galaxy 
($\epsilon$=0.28), and is only moderately concentrated (Harris et al. 1976; 
Djorgovski 1993).  Interestingly, many of these physical characteristics are
also observed in $\omega$ Cen.  Therefore, using the spectroscopic data 
presented here we aim to investigate the presence of a metallicity dispersion 
in NGC 6273, and to compare the cluster's light and heavy element abundance 
patterns with other clusters for which a spread in both [Fe/H] and s--process 
elements is confirmed.

\section{OBSERVATIONS, TARGET SELECTION, AND DATA REDUCTION}

\subsection{Observations and Target Selection}

NGC 6273 resides at a Galactocentric distance (R$_{\rm GC}$) of 0.7 kpc and a 
height above the plane of 1.4 kpc (e.g., Casetti--Dinescu et al. 2010).  
Therefore, the cluster is a member of the inner Galaxy globular cluster 
population, and possibly the kinematically hot inner halo or bulge.  Although 
NGC 6273 is only at a distance of $\sim$9 kpc from the Sun (e.g., Piotto et al.
1999), the large and highly variable reddening complicate both cluster RGB 
target selection and the interpretation of its color--magnitude diagram.  
Various estimates provide a color--excess of E(B--V)=0.31--0.47 and 
$\Delta$E(B--V)$\sim$0.2--0.3 (Racine 1973; Harris et al. 1976; Piotto et al. 
1999; Davidge 2000; Valenti et al. 2007; Brown et al. 2010; 
Alonso--Garc{\'{\i}}a et al. 2012).  

\begin{figure}
\epsscale{1.00}
\plotone{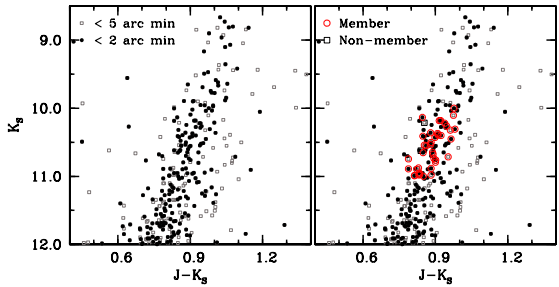}
\caption{\emph{left:} A K$_{\rm S}$ versus J--K$_{\rm S}$ color--magnitude
diagram, based on 2MASS data (Skrutskie et al. 2006), is shown for the stars
inside 2$\arcmin$ (filled black circles) and 5$\arcmin$ (open grey boxes)
of the cluster center.  \emph{right:} A plot similar to the left panel but
with stars identified as high probability radial velocity members (39 stars)
by open red circles and non--members (2 stars) indicated by open black boxes.
Note that one target identified as a non--member (2MASS 17023847--2618509)
exhibits a double--lined spectrum with one star at a radial velocity
consistent with cluster membership and the second star at much lower velocity.}
\label{f1}
\end{figure}

The differential reddening and contamination from outer bulge field stars can 
each broaden the RGB, and we are not aware of any dedicated membership studies
that are available for NGC 6273.  However, a selection of high probability 
cluster members can be made by examining the color--magnitude diagram at 
various radii.  As can be seen in the 2MASS (Skrutskie et al. 2006) 
color--magnitude diagram shown in Figure \ref{f1}, stars inside 2$\arcmin$ 
produce a broadened but well--defined RGB compared to those between 
2--5$\arcmin$.  Therefore, we selected stars approximately 1--2 magnitudes 
below the RGB--tip that reside inside 5$\arcmin$ from the cluster center and 
that lie along the sequence defined by stars inside 2$\arcmin$.  

Using the 2MASS astrometry and the Michigan/Magellan Fiber System (M2FS; Mateo 
et al. 2012) mounted on the Magellan--Clay 6.5m telescope at Las Campanas 
Observatory, we were able to deploy 47 fibers on one plug plate (41 stars/6 
sky).  The M2FS fiber system and MSpec spectrograph were configured using the 
high resolution ``Bulge$\_$GC1" setup described in Johnson et al. (2015), 
which produces cross--dispersed spectra spanning approximately 6120--6720 \AA. 
The 1.2$\arcsec$ fibers and 125$\mu$m slit provided a resolving power of
R$\equiv$$\lambda$/$\Delta$$\lambda$$\approx$27,000.  The spectra were obtained
from a single set of 3$\times$1800 sec exposures on 2014 June 2 under good 
seeing conditions (FWHM$\approx$0.5$\arcsec$), and this data set yielded 39/41 
(95$\%$)\footnote{Note that one of the two radial velocity non--members 
(2MASS 17023841--2618509) exhibits a double--lined spectrum with one star 
having a velocity consistent with cluster membership and the other star at a 
much lower velocity.  This is likely the result of two unrelated stars falling 
on the same fiber.} stars with radial velocities consistent with cluster 
membership (see Section 4 for more details).  The radial velocity member and 
non--member stars are identified on the color--magnitude diagram presented in 
Figure \ref{f1}.

\subsection{Data Reduction}

The data reduction process followed the same procedure outlined in Johnson
et al. (2015; see their Section 2.3).  To briefly summarize, the individual
amplifier images (four per CCD) of each exposure were bias subtracted and 
trimmed separately using the IRAF\footnote{IRAF is distributed by the National 
Optical Astronomy Observatory, which is operated by the Association of 
Universities for Research in Astronomy, Inc., under cooperative agreement with 
the National Science Foundation.} task \emph{ccdproc}.  Each set of four images
was then rotated and translated into the proper orientation and then combined
using the IRAF tasks \emph{imtranspose} and \emph{imjoin}.  The multi--fiber
reduction tasks such as aperture identification and tracing, scattered light
removal, flat--field correction, ThAr wavelength calibration, cosmic--ray
removal, and spectrum extraction, were carried out using repeated calls of
the IRAF task \emph{dohydra}.  The final reduction tasks including sky
subtraction, continuum fitting, spectrum combination, and telluric removal
were accomplished using the IRAF tasks \emph{skysub}, \emph{continuum},
\emph{scombine}, and \emph{telluric}.  The final combined spectra yielded
S/N of about 30--50 per resolution element.

\section{DATA ANALYSIS}

\subsection{Model Atmospheres}

Due to the presence of significant differential reddening across the cluster 
(e.g., see Alonso--Garc{\'{\i}}a et al. 2012; their Figure 13), we determined 
the model atmosphere parameters effective temperature (T$_{\rm eff}$), surface 
gravity (log(g)), metallicity ([Fe/H]), and microturbulence 
($\xi$$_{\rm mic.}$) using purely spectroscopic methods.  We performed an 
iterative process that solved for all four model atmosphere parameters 
simultaneously.  Specifically, the effective temperatures were derived by 
removing trends in log $\epsilon$(Fe I) as a function of excitation potential, 
the surface gravities were determined by forcing ionization equilibrium between
log $\epsilon$(Fe I) and log $\epsilon$(Fe II), and the microturbulences were 
determined by removing trends between log $\epsilon$(Fe I) and reduced 
equivalent width (EW)\footnote{The reduced equivalent width is defined as 
log(EW/$\lambda$).}.  The model atmosphere metallicity was set equal to the 
average [Fe/H] value determined from each iteration.  Since most stars in our 
sample have [$\alpha$/Fe]$\approx$$+$0.3, as measured from the average of our
derived [Mg/Fe], [Si/Fe], and [Ca/Fe] abundances (see Section 5), we 
interpolated within the available grid of $\alpha$--rich ATLAS9 model 
atmospheres provided by Castelli \& Kurucz (2004)\footnote{The model atmosphere
grid can be accessed at: http://wwwuser.oats.inaf.it/castelli/grids.html.}.  
However, the most metal--rich star in our sample (2MASS 17024453$-$2616377) has 
[$\alpha$/Fe]$\approx$0 so we used the scaled--solar grid for this case.
The final adopted model atmosphere parameters are provided in Table 1.

We note that the model atmosphere parameters adopted for this project do not 
include corrections due to possible departures from local thermodynamic 
equilibrium (LTE).  However, the expected effects on T$_{\rm eff}$, log(g), 
and $\xi$$_{\rm mic.}$ derived using a spectroscopic analysis and 1D LTE 
compared to 1D non--LTE are predicted to be $\la$50 K, $\la$0.10 (cgs), and 
$\la$0.10 km s$^{\rm -1}$, respectively, for the parameter space spanned here 
(Lind et al. 2012).  These differences are comparable to the internal precision
of our measurements.  A similar but possibly relevant issue related to 
differences in [Fe/H] derived from 1D LTE analyses of RGB, as opposed to AGB, 
stars has been noted by Ivans et al. (2001; NGC 5904), Lapenna et al. (2014; 
NGC 104), and Mucciarelli et al. (2015; NGC 3201).  These authors found that
the derived [Fe I/H] abundances were $\sim$0.10--0.15 dex lower than the 
[Fe II/H] abundances for AGB, but not RGB, stars when the model atmosphere
parameters were determined via spectroscopic methods.  Thus, mixing RGB and AGB
stars in the same sample could produce an artificial metallicity spread.  
Unfortunately, the large differential reddening and poor color separation of 
RGB and AGB stars in Figure \ref{f1} makes the assignment of our target stars 
to either evolutionary state difficult.  However, the short evolutionary time 
scale of AGB stars at the luminosity level reached by our observations suggests
a large fraction of the targets are likely first ascent red giants.  Additional 
evidence supporting the detection of a true metallicity spread is provided 
in Section 5.

\subsection{Abundance Analysis}

\subsubsection{Equivalent Width Measurements}

The abundances for all elements were derived using the \emph{abfind} and
\emph{synth} drivers of the LTE line analysis code MOOG\footnote{MOOG can be 
downloaded from http://www.as.utexas.edu/$\sim$chris/moog.html.} (Sneden 1973;
2014 version).  The [Fe/H], [Si/Fe], [Ca/Fe], [Cr/Fe], and [Ni/Fe] abundance
ratios were determined using EW measurements made with the semi--automated 
Gaussian profile fitting code developed for Johnson et al. (2014).  In 
general, we avoided measuring lines that were heavily blended, located near 
strong telluric features, or that had log(EW/$\lambda$)$\ga$--4.5.

The line selection, atomic transition parameters, and adopted solar abundances 
are included in Table 2.  The log(gf) values were determined through an 
inverse solar analysis by measuring the EWs of the lines from Table 2 in a 
daylight solar spectrum taken with the same M2FS configuration as the NGC 
6273 data.  However, in order to test for possible analysis differences 
between dwarf and giant stars, we also measured the same lines in the Arcturus 
atlas (Hinkle et al. 2000).  We adopted the model atmosphere parameters given
in Ram{\'{\i}}rez \& Allende Prieto (2011) and recovered similar abundances,
including the systematic offset of [Fe II/H]--[Fe I/H]$\approx$0.10 dex.
Therefore, we increased the solar--based Fe II log(gf) values by 0.10 dex to 
satisfy ionization equilibrium, and these changes are reflected in Table 2.
The final abundances of [Fe/H], [Si/Fe], [Ca/Fe], [Cr/Fe], and [Ni/Fe] are 
provided in Tables 3a--3b.

\subsubsection{Spectrum Synthesis Measurements}

The abundances of [Na/Fe], [Mg/Fe], [Al/Fe], [La/Fe], and [Eu/Fe] were measured
via spectrum synthesis rather than EW analyses because these elements provide
only a small number of often blended lines in the 6120--6720 \AA\ window used
here.  Additionally, the Mg line profiles near 6318 \AA\ can be affected by a 
broad Ca autoionization feature\footnote{Since our target stars are relatively
metal--poor, the Ca autoionization feature produced a $<$10$\%$ effect on the
Mg line profiles.}, and those of La and Eu are often broadened due to isotopic 
splitting and/or hyperfine structure.  Although the Na, Mg, and Al log(gf) 
values were determined using a similar procedure described in Section 3.2.1, 
we adopted the line lists and solar abundances of Lawler et al. (2001a,b) for 
La and Eu.  We also adopted the Solar System isotopic ratio of 47.8$\%$ for 
\iso{151}{Eu} and 52.2$\%$ for \iso{153}{Eu}.  Contamination by CN was 
accounted for in our syntheses by including in our line list the recently 
updated \iso{12}{C}\iso{14}{N} and \iso{13}{C}\iso{14}{N} line lists from 
Sneden et al. (2014).  For all stars we set [C/Fe]=--0.3, 
\iso{12}{C}/\iso{13}{C}=4, and estimated [O/Fe] based on the [Na/Fe] 
abundances.  In particular, the [O/Fe] abundances were estimated from a 
fit to the [O/Fe]--[Na/Fe] relation provided by Carretta et al. (2009c; their 
Figure 6) for several Galactic globular clusters.  The [N/Fe] abundance was 
adjusted as a free parameter to provide the best fit to nearby CN features.  
The final abundances of [Na/Fe], [Mg/Fe], [Al/Fe], [La/Fe], and [Eu/Fe] are 
provided in Tables 3a--3b.

\subsubsection{Internal Abundance Uncertainties}

The primary source of error for internal measurement precision comes from
uncertainties in the model atmosphere parameter determinations, with additional
contributions from line blending, continuum placement, atomic parameters, and 
visual profile fitting uncertainty.  The latter contributions are typically 
minor for reasonably high resolution and high S/N data, and we have estimated 
this (largely random) contribution by using the error of the mean for all 
species analyzed here.  On average, the abundance uncertainty on [X/H] ratios 
from measurement errors alone ranges from $\sim$0.02--0.05 dex.  

For the former error source, we estimate that the internal uncertainty for
T$_{\rm eff}$, log(g), [M/H], and $\xi$$_{\rm mic.}$ when constrained by
spectroscopic methods is approximately 50 K, 0.10 cgs, 0.07 dex, and 0.10
km s$^{\rm -1}$, respectively.  The T$_{\rm eff}$ and $\xi$$_{\rm mic.}$
estimates are derived from the scatter observed when fitting a linear function 
through plots of log $\epsilon$(Fe I) versus excitation potential (for 
T$_{\rm eff}$) and reduced equivalent width (for $\xi$$_{\rm mic.}$).  The 
uncertainty for surface gravity is estimated from the scatter in derived 
log(g) shown in Table 1 for stars within $\sim$100 K and with similar [Fe/H].
The model atmosphere metallicity uncertainty is estimated from the
measurement uncertainty of [Fe I/H] and [Fe II/H].  The total internal 
uncertainty for each element was calculated by adding in quadrature the 
measurement error and the change in abundance when each model atmosphere 
parameter was varied individually.  Note that for element ratios normalized by 
[Fe/H], the change in [Fe I/H] or [Fe II/H] was calculated simultaneously.  
This procedure ensures that the $\Delta$[X/Fe] uncertainties provided in 
Tables 3a--3b account for situations in which Fe and the element in question 
exhibit the same sign of variability when a given model atmosphere parameter 
is changed.

\section{RADIAL VELOCITY MEASUREMENTS AND CLUSTER MEMBERSHIP}

Radial velocities were determined using the \emph{fxcor} task in IRAF to cross 
correlate against the spectrum of Arcturus (Hinkle et al. 2000).  Heliocentric 
corrections were determined using the IRAF task \emph{rvcor}.  The spectral 
range from 6120--6275 \AA\ was used for cross correlation, avoiding strong 
telluric absorption features at wavelengths longer than 6275 \AA.  The Arcturus
spectrum was convolved with a Gaussian profile and rebinned to match the 
resolution and sampling of the observed spectra.  For the M2FS data, this 
spectral range is split across two orders, giving two independent measures of 
the radial velocity for each star and for each exposure.  The three exposures 
were measured individually, giving a total of six measurements for each star.  
The radial velocity dispersions provided in Tables 1 and 4 are the standard 
deviation of the six measurements.

\begin{figure}
\epsscale{1.00}
\plotone{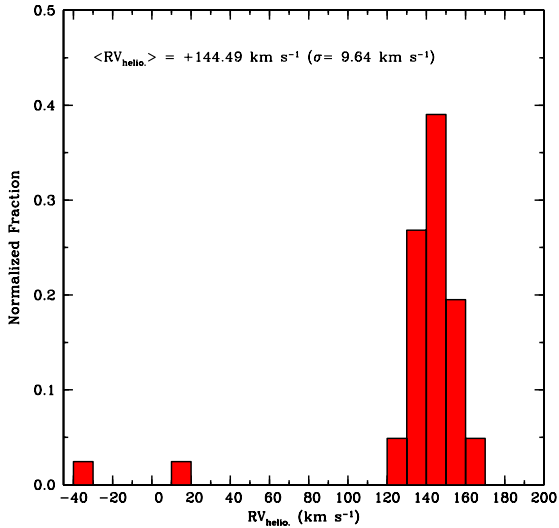}
\caption{A histogram showing the distribution of heliocentric radial velocity
(RV$_{\rm helio.}$) values measured for the stars shown in Figure \ref{f1}.
The data are sampled with 10 km s$^{\rm -1}$ bins.}
\label{f2}
\end{figure}

Although we were only able to derive abundances for 18/41 target stars, we 
measured radial velocities for all 41 RGB stars.  A histogram of our results
is shown in Figure \ref{f2}, and indicates that 39/41 (95$\%$) targets exhibit
radial velocities consistent with cluster membership.  Table 4 lists the 
coordinates, 2MASS photometry, and radial velocities for all observed stars 
not listed in Table 1.  We note that one star (2MASS 17023847--2618509) 
exhibits a double--lined spectrum with one component yielding a velocity 
consistent with cluster membership and the other component having a velocity 
inconsistent with cluster membership.  The remaining non--member star 
(2MASS 17024093--2620182) was determined to have [Fe/H]=--0.14, and is thus 
ruled out as a cluster member based on both kinematics and chemical composition.

For the stars that do exhibit velocities consistent with cluster membership,
we find an average heliocentric radial velocity of $+$144.49 km s$^{\rm -1}$
and a dispersion of 9.64 km s$^{\rm -1}$.  Our derived cluster velocity is
larger than the $+$135 km s$^{\rm -1}$ value given in Harris (1996; 2010 
version), and is also $\sim$20 km s$^{\rm -1}$ larger than the 
$+$121 km s$^{\rm -1}$ ($\sigma$=10.6 km s$^{\rm -1}$) value provided by
Webbink (1981).  Additional literature systemic velocities range from $+$120
km s$^{\rm -1}$ to $+$147 km s$^{\rm -1}$, but there is general agreement that
the internal dispersion is $\sim$10 km s$^{\rm -1}$ (Zinn \& West 1984; Hesser 
et al. 1986; Rutledge et al. 1997).  It is possible that the large range in 
derived cluster velocities stems from Galactic bulge field star contamination, 
which can be significant if observations extend beyond a few arc minutes from 
the cluster core.

Although NGC 6273 lies near the crowded Galactic bulge 
(\emph{l}=--3.1,\emph{b}=$+$9.4), we do not believe a significant fraction of
the stars residing near the cluster's average radial velocity are contaminating
bulge field stars.  First, the average Galactocentric radial velocity
of bulge field stars near \emph{l}=--3.1 should be approximately --50 
km s$^{\rm -1}$ with a dispersion of about 80--100 km s$^{\rm -1}$ (e.g.,
Kunder et al. 2012; Ness et al. 2013a; Zoccali et al. 2014).  The expected
bulge field velocities are in stark contrast to the average Galactocentric 
velocity of our claimed cluster members at $+$142.02 km s$^{\rm -1}$ 
($\sigma$=9.64 km s$^{\rm -1}$).  Second, the Galactic bulge metallicity
distribution function cuts off considerably at [Fe/H]$\la$--1 (e.g., 
Zoccali et al. 2008; Bensby et al. 2013; Johnson et al. 2013; Ness et al. 
2013b), and as will be discussed in Section 5.1 the [Fe/H] range of our claimed 
cluster members spans --1.80 to --1.30.  Therefore, finding bulge field stars 
with [Fe/H]$\sim$--1.5, a high positive velocity, and projected near 
NGC 6273 should be an extremely rare event.

\section{ABUNDANCE RESULTS}

\subsection{Evidence Supporting a Metallicity Spread}

Previous analyses of NGC 6273 have hinted that the cluster may possess an 
intrinsic metallicity dispersion, as evidenced by a potentially broadened RGB 
(e.g., Harris et al. 1976; Piotto et al. 1999) and spread in the near--infrared
Ca II triplet equivalent width measurements by Rutledge et al. (1997).  As 
mentioned in Section 1, NGC 6273 even shares several observational 
characteristics (e.g., blue HB; elliptical morphology) with the chemically 
diverse $\omega$ Cen.  However, further interpretation linking these results 
to an intrinsic metallicity spread have been hindered by the cluster's 
significant differential reddening and location near the crowded, and largely 
metal--rich, Galactic bulge.  

\begin{figure}
\epsscale{1.00}
\plotone{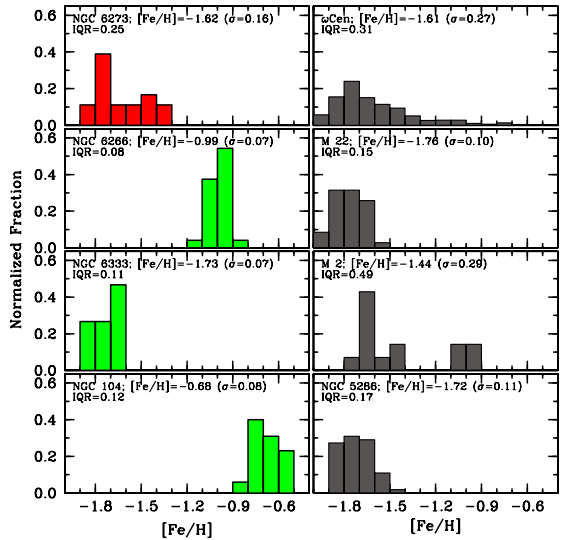}
\caption{\emph{left:} Metallicity distribution functions are shown for NGC 6273
and three other globular clusters observed with M2FS.  The results for NGC 6266
and 6333 will be presented in a future publication, but were observed during
the same observing run as NGC 6273.  The NGC 104 data are available in Johnson
et al. (2015). \emph{right:} Metallicity distribution functions are shown for
four additional globular clusters with confirmed metallicity spreads.  The
data for $\omega$ Cen, M 22, M 2, and NGC 5286 are from: Johnson \& Pilachowski
(2010), Marino et al. (2011b), Yong et al. (2014), and Marino et al. (2015),
respectively.  The average [Fe/H] value, dispersion, and interquartile range
are provided in all panels, and all data are sampled in 0.1 dex bins.}
\label{f3}
\end{figure}

The data presented here provide the first detailed high resolution 
spectroscopic analysis of cluster RGB members, and we find several lines of
evidence supporting the presence of an intrinsic metallicity dispersion within
NGC 6273.  In Figure \ref{f3} we show binned metallicity distribution functions
for NGC 6273, three monometallic clusters analyzed with M2FS data from the same 
observing run (NGC 104, NGC 6266, and NGC 6333), and four iron--complex 
clusters from the literature with confirmed metallicity spreads ($\omega$ Cen,
M 2, M 22, and NGC 5286).  The three monometallic clusters have [Fe/H] 
dispersions and interquartile ranges (IQR) of approximately 0.07 and 0.10 dex, 
respectively.  However, using the same instrument and setup, we find NGC 6273 
to have $\sigma$$_{\rm [Fe/H]}$=0.16 dex and IQR$_{\rm [Fe/H]}$=0.25 dex.  The
larger [Fe/H] dispersion and IQR for NGC 6273 is more in--line with 
observations of the iron--complex clusters, which have 
$\sigma$$_{\rm [Fe/H]}$$\geq$0.10 dex and IQR$_{\rm [Fe/H]}$$\geq$0.15 dex,
respectively, when [Fe/H] is derived from high resolution, high S/N spectra. 

The overall cluster average of [Fe/H]=--1.62 derived here for NGC 6273 is 
consistent with previous estimates from photometry (Davidge 2000; Valenti et 
al. 2007) and calibrated spectroscopic indices (Zinn \& West 1984; Kraft \& 
Ivans 2003; Carretta et al. 2009a), which range from [Fe/H]=--1.9 to --1.4.
The [Fe/H] distribution shown in Figure \ref{f3} exhibits evidence of at least 
two distinct stellar populations with different metallicities, including a 
possible third more metal--rich population that has [$\alpha$/Fe]$\approx$0.  
Therefore, we have separated the stars into three groups, based on each star's 
[Fe/H] abundance, which have $\langle$[Fe/H]$\rangle$=--1.75 ($\sigma$=0.04; 
``metal--poor"), $\langle$[Fe/H]$\rangle$=--1.51 ($\sigma$=0.08; 
``metal--rich"), and $\langle$[Fe/H]$\rangle$=--1.30 (1 star; ``anomalous").  
We find that the metal--poor population dominates by number (50$\%$) compared 
to the metal--rich (44$\%$) and anomalous (6$\%$) groups.  However, since
we did not target the reddest stars in Figure \ref{f1}, the fraction of 
cluster stars with [Fe/H]$\ga$--1.30 may be higher than the 6$\%$ measured 
here.  Although the total sample size is only 18 stars, the average 
heliocentric radial velocities and dispersions are similar between the 
metal--poor, metal--rich, and anomalous groups (see Table 5).  The similar 
velocities strongly suggest that all three sub--populations are members of NGC 
6273. 
 
\begin{figure}
\epsscale{1.00}
\plotone{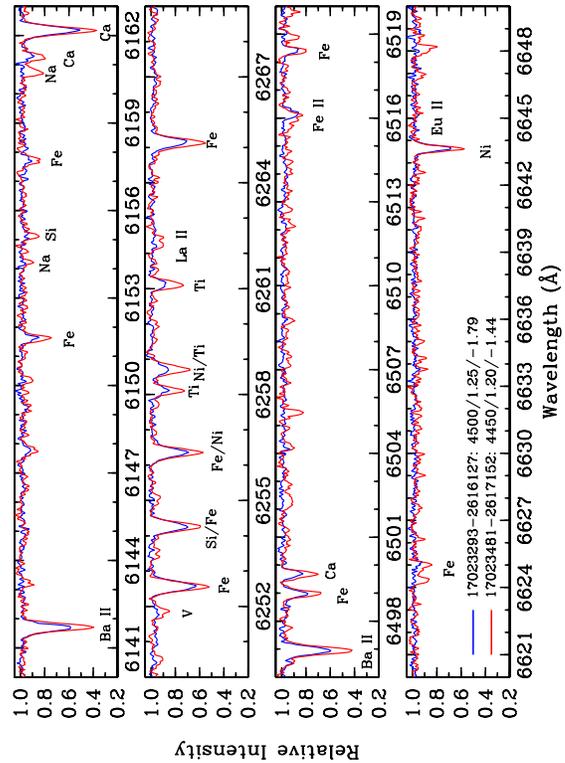}
\caption{Sample spectra are shown comparing the line strengths of two stars
with similar T$_{\rm eff}$ and log(g) but different metallicity.  The blue
spectrum (2MASS 17023293--2616127) is a sample from the metal--poor population
and the red spectrum (2MASS 17023480--2617152) is a sample from the metal--rich
population.}
\label{f4}
\end{figure}

Further evidence in support of a metallicity spread can be seen by a 
visual examination of the spectra.  In Figure \ref{f4} we compare the M2FS 
spectra of two stars with similar T$_{\rm eff}$ and log(g) but that differ
in [Fe/H] by more than a factor of two.  Note that the line strengths are
greater for nearly all transitions in the metal--rich star.  The significant 
increase in EW for La II and Ba II is especially noteworthy.  In the context 
of similar chemical patterns being present in other iron--complex clusters (see 
Section 6), the observed correlation between [Fe/H] and [La/Fe] in NGC 6273 
offers compelling evidence that the [Fe/H] dispersion is real.

\begin{figure}
\epsscale{1.00}
\plotone{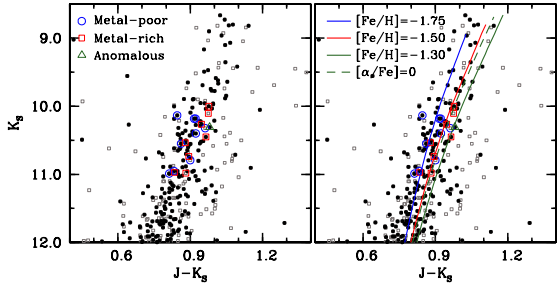}
\caption{Color--magnitude diagrams similar to Figure \ref{f1} are shown but
with the observed high probability cluster members designated by metallicity.
The metal--poor ([Fe/H]$\approx$--1.75), metal--rich ([Fe/H]$\approx$--1.50),
and ``anomalous" ([Fe/H]$\approx$--1.30) populations are designated with the
open blue circles, open red boxes, and open green triangles, respectively.
Dartmouth stellar isochrones (Dotter et al. 2008) with ages of 12 Gyr,
[Fe/H]=--1.75, --1.50, and --1.30, and [$\alpha$/Fe]=$+$0.20 are overlaid as
blue, red, and green solid lines, respectively.  We assumed a distance of
9 kpc (Piotto et al. 1999) and E(B-V)=0.30, which provided the best agreement
between photometric and spectroscopic temperatures (see Section 5.1).  A
similar isochrone with [Fe/H]=--1.30 and [$\alpha$/Fe]=0, which matches more
closely to the abundance pattern of the anomalous star, is plotted as a dashed
green line.}
\label{f5}
\end{figure}

\begin{figure}
\epsscale{1.00}
\plotone{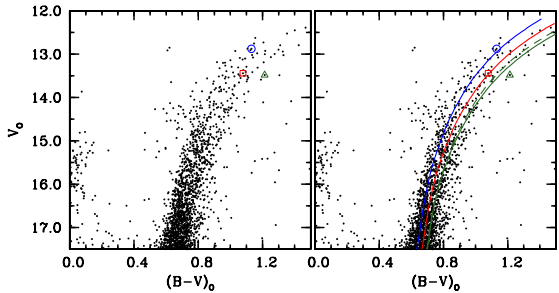}
\caption{Hubble Space Telescope (HST) optical color--magnitude diagrams are
shown for the RGB regions of NGC 6273.  The data are from Piotto et al. (1999;
2002) and have been dereddened following the prescriptions outlined in the
original papers.  The right panel shows the same color--magnitude diagrams but
with the isochrones from Figure \ref{f5} over plotted.  The open symbols are
the same as in Figure \ref{f5}, and represent three stars (2MASS
17023868--2616516, 17024016--2616096, and 17024453--2616377) we were able to
match from our sample onto the Piotto et al. (1999) HST images.}
\label{f6}
\end{figure}

A similar argument is made by examining the location of stars on the 
color--magnitude diagram when separated by [Fe/H].  Figure \ref{f5} shows that,
even in a near--infrared color--magnitude diagram that is not corrected for
differential reddening, for a given luminosity level the more metal--poor stars
tend to be bluer than the more metal--rich stars.  Furthermore, the RGB 
color dispersion may be largely explained by the presence of differential
reddening\footnote{We examined the validity of both the absolute and 
differential reddening values noted in Section 2 by using our spectroscopically
derived temperatures and 2MASS J--K$_{\rm S}$ colors to invert the
color--temperature relation provided by Gonz{\'a}lez Hern{\'a}ndez \& 
Bonifacio (2009).  Assuming E(J--K$_{\rm S}$)/E(B--V)=0.505 (Fiorucci \& Munari
2003), we determined the E(B--V) value for each program star that enabled a 
match between the photometric and spectroscopic T$_{\rm eff}$ values.  We 
determined the best--fit average E(B--V)=0.30 mag. with a standard deviation 
of 0.11 mag and a full range of about 0.30 mag.  These values are in reasonable
agreement with the previous estimates mentioned in Section 2.} and a $\sim$0.5 
dex spread in [Fe/H].  A similar conclusion is reached when examining the 
dereddened optical color--magnitude diagram from Piotto et al. (1999; 2002) 
shown in Figure \ref{f6}.  We were able to match one star from each 
spectroscopically defined sub--population to the original Hubble Space 
Telescope images from Piotto et al. (1999; 2002), and again we find that the 
stars are distributed as one would expect if the RGB color dispersion was at 
least partially driven by a metallicity spread.  Interestingly, both 
Figures \ref{f5} and \ref{f6} show evidence of stars residing at even redder 
colors than were observed here.  While these may belong to the Galactic bulge 
field population, they are found projected near the dense cluster core.  If 
some fraction of these presumably more metal--rich stars are cluster members, 
then the metallicity spread would exceed 0.5 dex.  Such a large [Fe/H] spread 
would make NGC 6273 more similar to $\omega$ Cen and M 2 than the less extreme 
iron--complex clusters.

\subsection{Light Odd--Z and $\alpha$ Element Abundances}

\begin{figure}
\epsscale{1.00}
\plotone{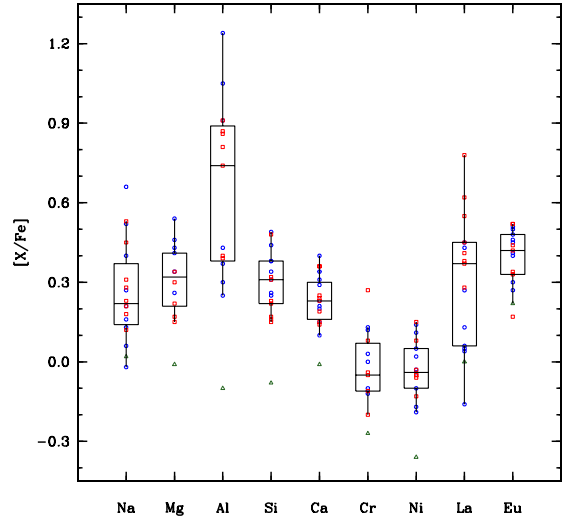}
\caption{A box plot illustrating the [X/Fe] ratios of all elements measured
here.  The target stars are designated using the same colors and symbols as
in Figure \ref{f5}.  For each element column the three horizontal lines
indicate the first quartile, median, and third quartile values for all stars.
The vertical lines indicate the full range of [X/Fe] values, excluding outliers
($>$1.5$\times$ the interquartile range).}
\label{f7}
\end{figure}

\begin{figure}
\epsscale{1.00}
\plotone{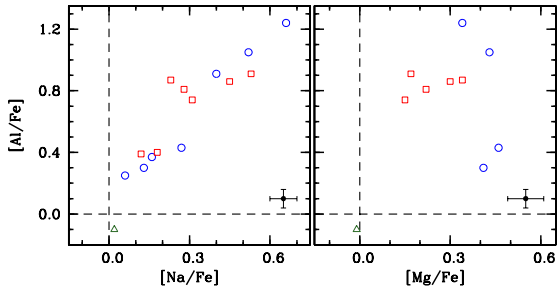}
\caption{\emph{left:} A plot of [Al/Fe] versus [Na/Fe] is shown for NGC 6273
stars.  Similar to Figure \ref{f5}, the metal--poor, metal--rich, and
``anomalous" RGB stars are designated as open blue circles, open red boxes,
and open green triangles.  Note that both main populations exhibit the
well--known Na--Al correlation characteristic of high--temperature
proton--capture nucleosynthesis, and also exhibit a noticeable gap between
$+$0.45$\la$[Al/Fe]$\la$$+$0.75.  The dashed lines in both panels represent the
solar abundance ratios.  \emph{right:} A plot of [Al/Fe] versus [Mg/Fe] for
the same stars as in the left panel.  The sample size of stars within each
sub--population is too small to conclude whether or not a Mg--Al
anti--correlation is present.  Typical error bars are shown (black symbol) in
the bottom right corner of each panel.}
\label{f8}
\end{figure}

In Figure \ref{f7} we present a box plot of the [X/Fe] ratios for all elements
analyzed here.  For the elements with Z$\leq$20, [Mg/Fe] ($\sigma$=0.12 dex), 
[Si/Fe] ($\sigma$=0.11 dex), and [Ca/Fe] ($\sigma$=0.09 dex) have the smallest
abundance dispersions.  From these data and the abundance uncertainties listed
in Tables 3a--3b, we conclude that a dispersion of $\sim$0.10 dex is the 
limit separating elements with and without significant abundance spreads.
Therefore, among the light elements both [Na/Fe] ($\sigma$=0.18 dex) and 
[Al/Fe] ($\sigma$=0.31 dex) exhibit substantial star--to--star scatter.  
However, the abundances of [Na/Fe] and [Al/Fe] are correlated (see 
Figure \ref{f8}), and [Al/Fe] even shows some evidence of a bimodal 
distribution.  In particular, we do not find any stars with 
$+$0.45$\la$[Al/Fe]$\la$$+$0.75, and the gap appears to be present in both the
metal--poor and metal--rich groups.  We note that a similar feature has been
observed in other clusters, such as $\omega$ Cen (Norris \& Da Costa 
1995; Johnson et al. 2008; Johnson \& Pilachowski 2010) and NGC 2808 (Carretta 
2014), as well.  Figure \ref{f8} shows some (weak) evidence in support of a
Mg--Al anti--correlation, which is expected if Al production is driven by
the MgAl proton--capture cycle.  However, more data are needed to confirm
this result.

Although both the metal--poor and metal--rich populations independently show
a Na--Al correlation, Figures \ref{f7}--\ref{f8} and Table 5 indicate that
the dispersion is not equivalent between the two main sub--populations.  The
metal--poor and metal--rich groups have similar average [Na/Fe] and [Al/Fe]
abundances, but the star--to--star dispersion is larger for the metal--poor 
group.  However, the dispersion for [Mg/Fe], [Si/Fe], and [Ca/Fe] is 
essentially identical between the metal--poor and metal--rich groups.  The
metal--poor stars have slightly higher [Mg/Fe], [Si/Fe], and [Ca/Fe] abundances
than the metal--rich stars, but both populations are $\alpha$--enhanced
([$\alpha$/Fe]$\approx$$+$0.30).  In fact, the average difference in overall 
[$\alpha$/Fe] between the two populations is only 0.06 dex. 

The single anomalous star in our sample is the most metal--rich 
([Fe/H]=--1.30) and also has a peculiar composition.  Unlike the metal--poor
and metal--rich groups, the anomalous star has [X/Fe]$\sim$0 for all elements
with Z$\leq$20 measured here.  As can be seen in Figure \ref{f7}, the anomalous
star's abundance pattern is an outlier for all light elements except [Na/Fe].
Although it is not unusual for a globular cluster to host stars with 
[Na/Fe]$\sim$0 and/or [Al/Fe]$\sim$0, very few clusters have 
[$\alpha$/Fe]$\sim$0 (e.g., see Gratton et al. 2004; their Figure 4).  
Interestingly, the few clusters confirmed to have low mean [$\alpha$/Fe] 
abundances, such as Rup 106 and Pal 12 (Brown et al. 1997; Cohen 2004), may 
not be native to the Milky Way and are proposed to be relics of captured 
systems (e.g., Lin \& Richer 1992; Law \& Majewski 2010).  The anomalous 
star's radial velocity and location on the color--magnitude diagram support
the notion that it is a cluster member.  However, one study cannot 
definitively constrain whether the anomalous star was formed \emph{in situ} 
with the metal--poor and/or metal--rich groups, or was captured from another 
system or stream.

\subsection{Fe--Peak Element Abundances}

As is evident from Figure \ref{f7}, both Cr ($\langle$[Cr/Fe]$\rangle$=--0.01 
dex; $\sigma$=0.13 dex) and Ni ($\langle$[Ni/Fe]$\rangle$=--0.02 dex; 
$\sigma$=0.10 dex) closely track Fe and have approximately the same average
abundance\footnote{The anomalous star is excluded from the average and standard
deviation calculations here.}.  Similarly, both the metal--poor and metal--rich
sub--populations have equivalent [Cr/Fe] and [Ni/Fe] abundances and 
dispersions (see Table 5).  The relatively small star--to--star scatter and 
[X/Fe]$\approx$0 pattern for the Fe--peak elements is typical of most Galactic 
globular clusters (e.g., see review by Gratton et al. 2004).

In addition to having [$\alpha$/Fe]$\sim$0, the anomalous star exhibits a 
very peculiar Fe--peak abundance pattern with [Cr/Fe]=--0.27 and 
[Ni/Fe]=--0.36.  The factor of two depletion for Cr and Ni relative to Fe is
particularly puzzling because both elements typically follow similar
production patterns with Fe.  We are only aware of a few cases in which the 
even--Z Fe--peak elements have a significant depletion relative to Fe, 
including the aforementioned Rup 106 and Pal 12 (Brown et al. 1997; Cohen 
2004\footnote{We note that Cohen (2004) derived low [Ni/Fe] but [Cr/Fe]$\sim$0
for Pal 12.}), Terzan 7 (Tautvai{\v s}ien{\.e} et al. 2004; Sbordone et al. 
2005), and the open cluster M 11 (Gonzalez \& Wallerstein 2000).  There is
also a small but growing number of individual stars with similar even--Z 
element deficiencies such as CS 22169--035 ([Fe/H]=--3.04; Cayrel et al. 2004),
Car--612 ([Fe/H]=--1.30; Venn et al. 2012), HE 1207--3108 ([Fe/H]=--2.70;
Yong et al. 2013), and the modestly metal--poor field stars HD 193901
and HD 194598 ([Fe/H]$\approx$--1.15; Jehin et al. 1999; Gratton et al. 2003; 
Jonsell et al. 2005).  Several of the systems observed to have low [Cr,Ni/Fe] 
are associated with the captured Sagittarius dwarf spheroidal system, and 
nearly all of the low [Ni/Fe] objects also have low [$\alpha$/Fe].  The similar
abundance pattern of these objects with the anomalous star in NGC 6273 suggests
that, assuming the anomalous star is a cluster member, the final stage of star 
formation in NGC 6273 may have proceeded under conditions not experienced by 
most globular clusters (e.g., unusual initial mass function; long time delay) 
or that NGC 6273 may have accreted stars or gas from another system.

\subsection{Neutron--Capture Element Abundances}

\begin{figure}
\epsscale{1.00}
\plotone{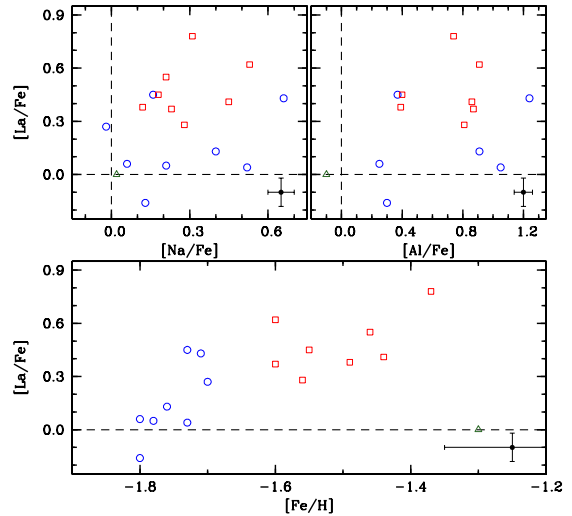}
\caption{The top two panels are similar to Figure \ref{f8} except [La/Fe] is
plotted as a function of [Na/Fe] (left) and [Al/Fe] (right).  The bottom panel
shows [La/Fe] plotted as a function of [Fe/H].  Note the strong rise in
[La/Fe] as a function of increasing [Fe/H].  A typical error bar is shown in
the bottom right corner of each panel.}
\label{f9}
\end{figure}

\begin{figure}
\epsscale{1.00}
\plotone{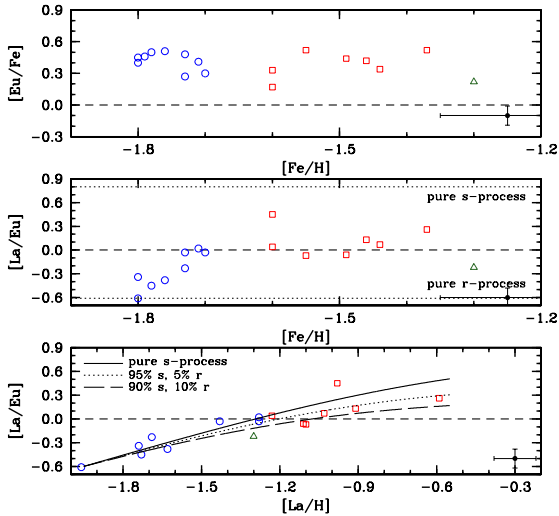}
\caption{The top and middle panels plot [Eu/Fe] and [La/Eu] as a function of
[Fe/H].  The [La/Eu] ratio is a measure of the relative contributions from the
s--process and r--process.  The increasing [La/Eu] ratio with increasing
metallicity is a characteristic of s--process enrichment.  Following McWilliam
et al. (2013), the bottom panel shows [La/Eu] as a function of [La/H].  The
solid black line illustrates the expected enrichment profile when pure
s--process material is added to a pure r--process composition.  The dotted and
long dashed lines indicate the expected profiles for 95$\%$(s)/5$\%$(r) and
90$\%$(s)/10$\%$(r) mixtures added to a pure r--process composition.  The
remaining colors and symbols are the same as those in Figure \ref{f5}, and
the pure r--process and s--process values are from Kappeler et al. (1989) and
Bisterzo et al. (2010), respectively.}
\label{f10}
\end{figure}

We find that the dispersion for [La/Fe] ($\sigma$=0.25 dex) is significantly 
larger than for [Eu/Fe] ($\sigma$=0.11 dex), and Figure \ref{f7} indicates that 
the [La/Fe] dispersion is only exceeded by that of [Al/Fe].  However, 
Figure \ref{f9} shows that [La/Fe] is not well correlated with either [Na/Fe] 
or [Al/Fe], which suggests that the two element groups are produced by 
different sources.  Interestingly, [La/Fe] is strongly correlated with [Fe/H] 
and increases steadily from [La/Fe]$\sim$--0.10 in the most metal--poor 
stars to [La/Fe]$\sim$$+$0.50 in the most metal--rich stars (except for the 
anomalous star).  In contrast, the Eu abundance remains relatively constant at 
[Eu/Fe]$\approx$$+$0.40 (see Figure \ref{f10}).  The strong rise in [La/Fe] as 
a function of increasing [Fe/H], coupled with the enhanced and nearly constant 
[Eu/Fe] abundance, is a trait that is common to many iron--complex globular 
clusters (see Section 6).

A comparison of the [La/Fe], [Eu/Fe], and [La/Eu] ratios as a function of 
sub--population, [Fe/H], and [La/H] is shown in Figures \ref{f9}--\ref{f10} 
and Table 5.  The [La/Fe] dispersion is marginally larger for the metal--poor 
population, and in fact [La/Fe] rises as a function of [Fe/H] approximately 
twice as fast in the metal--poor group as the metal--rich group.  However, the 
anomalous star deviates from the overall cluster trend and has [La/Fe]=0, 
despite being the most metal--rich star.  Since [Eu/Fe] is nearly constant
across all three sub--populations, the change in [La/Eu], which is a measure
of the relative contributions from the s--process and r--process and is
largely insensitive to model atmosphere uncertainties, with metallicity follows the [La/Fe] trend.  As can be seen in Figure \ref{f10}, [La/Eu] increases from 
[La/Eu]$\sim$--0.60 to [La/Eu]$\sim$0.00 in the metal--poor population.  The 
[La/Eu] ratio either remains constant at [La/Eu]$\sim$0.00 or slowly increases 
for the metal--rich population.  The anomalous star has [La/Eu]$\sim$--0.20, 
which is more in--line with some of the more metal--poor stars.

The peculiar shape of the [La/Eu] versus [Fe/H] distribution warrants further 
investigation.  The low [La/Eu] ratio for the most metal--poor stars in 
NGC 6273 is a characteristic shared with most globular clusters, and is
consistent with a pure r--process distribution.  Therefore, the rise in 
[La/Fe], but not [Eu/Fe], with increasing [Fe/H] suggests the [La/Eu] 
increase is driven primarily by s--process production.  Following McWilliam
et al. (2013) and adopting [La/Eu]=--0.60 (Kappeler et al. 1989) and $+$0.80 
(Bisterzo et al. 2010) as the pure r--process and s--process ratios, in 
Figure \ref{f10} we plot [La/Eu] as a function of [La/H].  We also include
three simple dilution models that mix pure s--process, 95$\%$ s--process, and
90$\%$ s--process material with an initial pure r--process composition.  As
noted by McWilliam et al. (2013), this plot removes the production degeneracy
of [Fe/H] since Fe can be produced in both Type II and Type Ia supernovae.  
The data and models in Figure \ref{f10} suggest that the metal--poor
population may be well--fit by assuming the dispersion in [La/Eu] is driven
by mixing pure s--process material with the initial (primordial?) pure 
r--process composition.  Interestingly, the more metal--rich stars are better
fit by a model that assumes a 5--10$\%$ contribution from the r--process, which
is perhaps an indication that the stars producing Fe for the metal--rich 
generation also synthesized some Eu.  The anomalous star is again peculiar in
composition and may be consistent with a model that assumes a $>$10$\%$ 
contribution from the r--process.  

\section{COMPARISON WITH OTHER ``IRON--COMPLEX" GLOBULAR CLUSTERS}

\begin{figure}
\epsscale{1.00}
\plotone{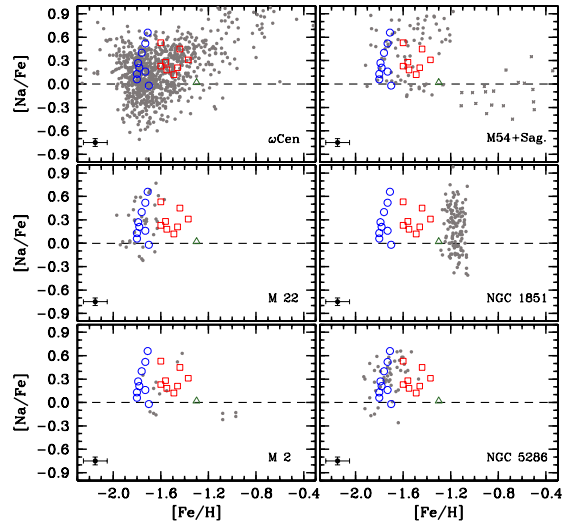}
\caption{[Na/Fe] abundances are plotted as a function of [Fe/H] for NGC 6273
(open blue circles, red boxes, and green triangles as in Figure \ref{f5}) and
six additional iron--complex globular clusters.  The data for $\omega$ Cen,
M 22, M 2, M54 $+$ Sagittarius, NGC 1851, and NGC 5286 are from: Carretta et
al. (2010a,b), Johnson \& Pilachowski (2010), Carretta et al. (2011), Marino et
al. (2011b), Yong et al. (2014), and Marino et al. (2015).  The dashed lines
indicate the solar abundance ratios.}
\label{f11}
\end{figure}

\begin{figure}
\epsscale{1.00}
\plotone{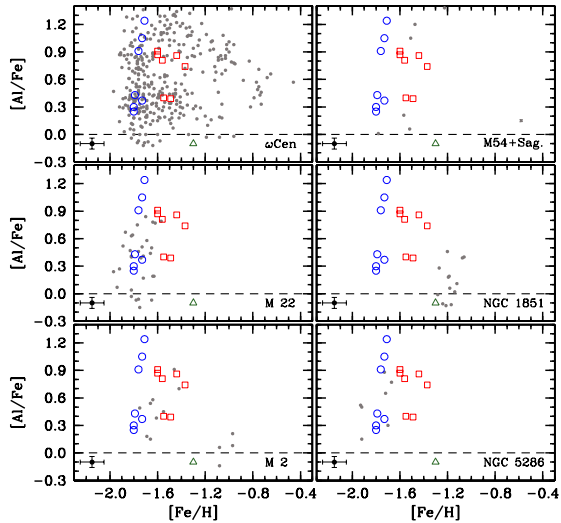}
\caption{A similar plot to Figure \ref{f11} showing [Al/Fe] as a function of
[Fe/H].}
\label{f12}
\end{figure}

\begin{figure}
\epsscale{1.00}
\plotone{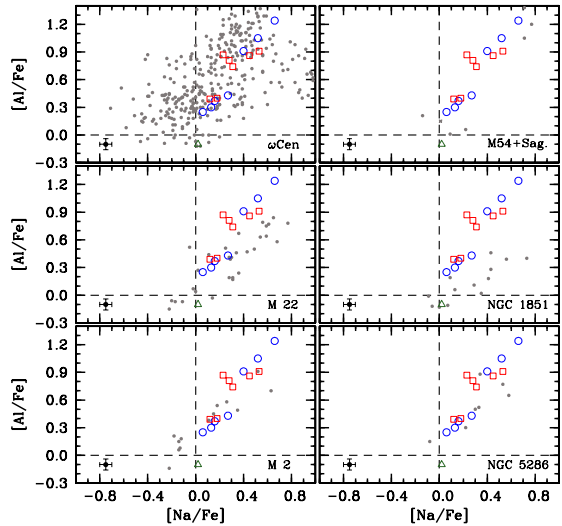}
\caption{A similar plot to Figure \ref{f11} showing [Al/Fe] as a function of
[Na/Fe].}
\label{f13}
\end{figure}

\begin{figure}
\epsscale{1.00}
\plotone{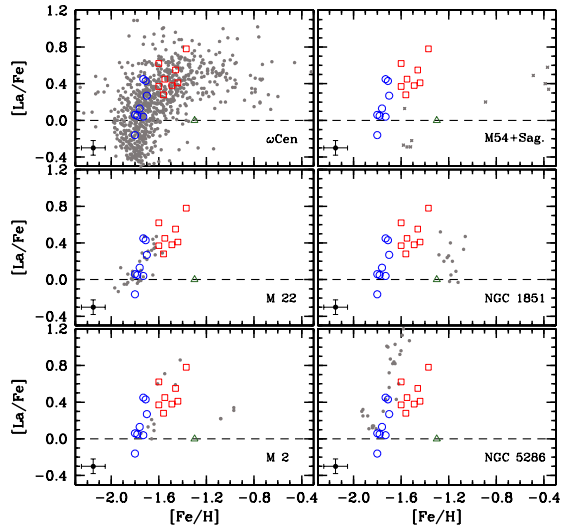}
\caption{A similar plot to Figure \ref{f11} showing [La/Fe] as a function of
[Fe/H].  Additional references for M54 $+$ Sagittarius include Brown et al.
(1999), Sbordone et al. (2007), and McWilliam et al. (2013).}
\label{f14}
\end{figure}

\begin{figure}
\epsscale{1.00}
\plotone{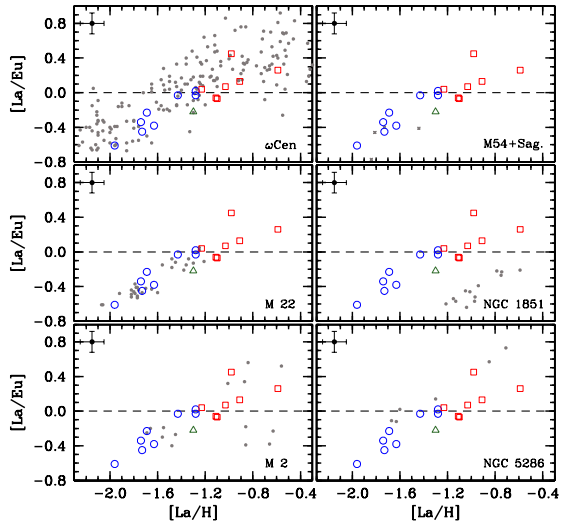}
\caption{A similar plot to Figures \ref{f10}--\ref{f11} showing [La/Eu] as a
function of [La/H].}
\label{f15}
\end{figure}

As mentioned in Section 1, the peculiar metallicity spread and heavy element 
abundance patterns exhibited by NGC 6273 separate the cluster from the bulk
of the Galaxy's globular cluster population.  Instead, NGC 6273 may belong to
a growing class of iron--complex clusters that still exhibit the large 
star--to--star light element abundance dispersions that define a system as a 
globular cluster, but that also have $\sigma$$_{\rm [Fe/H]}$$\ga$0.10, 
IQR$_{\rm [Fe/H]}$$\ga$0.15, and large s--process abundance spreads that are
correlated with [Fe/H].  A comparison between NGC 6273 and the iron--complex 
clusters $\omega$ Cen, M 22, M 2, M54$+$Sagittarius dwarf spheroidal galaxy,
NGC 1851, and NGC 5286 are summarized in Figures \ref{f11}--\ref{f15}.  Even
though all of these systems exhibit significant [Fe/H] and heavy element 
spreads and are included for context, we will primarily focus on comparing NGC 
6273 with $\omega$ Cen, M 22, M 2, and NGC 5286.  These four clusters appear 
to share the closest chemical composition and formation history with NGC 6273.

A comparison of the NGC 6273 [Na/Fe] abundances with the other iron--complex 
clusters is shown in Figures \ref{f11}.  In general, we find that NGC 6273 
shares a similar average [Na/Fe] abundance and star--to--star dispersion with
the other clusters.  Additionally, Figure \ref{f11} indicates that there may 
be a pattern of decreasing [Na/Fe] dispersion for the more metal--rich 
populations of NGC 6273, $\omega$ Cen, and possibly NGC 5286.  We note that a 
decrease in [Na/Fe] dispersion with increasing [Fe/H] would fit a global trend 
observed in monometallic globular clusters as well (e.g., Carretta et al. 
2009b, their Figure 3; Johnson \& Pilachowski 2010, their Figure 15).  A 
decrease in both the maximum [Al/Fe] abundance and dispersion for the
more metal--rich populations is clearly seen in both NGC 6273 and $\omega$ Cen
(Figure \ref{f12}), and also matches trends observed in monometallic clusters 
(e.g., Carretta et al. 2009b; O'Connell et al. 2011; Cordero et al. 2014, 
2015).  The prominent bimodal [Al/Fe] distribution of NGC 6273 is also clearly
seen in $\omega$ Cen, M54$+$Sagittarius, and possibly M 2.  Plotting [Al/Fe]
as a function of [Na/Fe] in Figure \ref{f13} further reveals that most of the
iron--complex clusters likely host discreet light element populations and share 
a common Na--Al correlation slope.  Since most of the clusters in 
Figure \ref{f13} have approximately the same metallicity (see also 
Figure \ref{f3}), the common Na--Al correlation slope may be an indication
that a similar mass range of sources polluted each cluster.

The discreet nature of the Na--Al correlation is not surprising given that both
normal and iron--complex clusters exhibit well--separated photometric sequences 
when observed with filters sensitive to light element abundances (e.g., Piotto 
et al. 2015).  However, since the light element (anti--)correlations are 
present in both normal clusters and in the various populations hosted by most
iron--complex clusters, it is clear that the process which produces a 
metallicity and neutron--capture abundance spread is independent of light 
element production.

Although nearly all globular clusters exhibit similar light element abundance
patterns, the correlated increase in [La/Fe] with [Fe/H] is perhaps the most
unusual and defining characteristic of iron--complex clusters.  As can be seen
in Figures \ref{f14}--\ref{f15}, the shape and slope of the [La/Fe] and [La/Eu]
distributions are nearly identical for all of the iron--complex clusters (see 
also Marino et al. 2015).  Combining the information from Figures \ref{f10} and
\ref{f15} indicates that in all of the iron--complex clusters included here the 
increase in [La/Eu] is due to almost pure s--process enrichment.  Previous
work on iron--complex clusters such as M 2 (Lardo et al. 2013; Yong et al. 
2014), M 22 (Marino et al. 2009, 2011b), and NGC 5286 (Marino et al. 2015) 
suggested that each may be decomposed into at least two populations: a low 
[La/Eu] metal--poor group and a high [La/Eu] metal--rich group.  The data 
presented in Figures \ref{f14}--\ref{f15} support these findings, and 
Figure \ref{f15} in particular suggests that clusters with similar [Fe/H] also
differentiate in [La/Eu] at about the same [La/H] abundance.  For clusters 
with [Fe/H]$\sim$--1.7, the increase in [La/Eu] occurs near [La/H]$\sim$--1.6
to --1.4.

The similar heavy element abundance patterns observed for many iron--complex 
clusters makes it tempting to speculate that all of them formed through the 
same basic process.  However, just as both normal and iron--complex clusters 
exhibit similar light element abundance variations but can have very different
heavy element distributions, the similar s--process abundances of the 
iron--complex clusters could be a red herring.  In other words, different 
formation mechanisms, timescales, and/or evolution paths may produce clusters 
with similar composition characteristics.  For example, M 22, M 2, and 
$\omega$ Cen exhibit a similar increase in [La/Eu] over the same [La/H] range, 
but the s--process production site may not be the same.  Roederer et al. 
(2011) and Yong et al. (2014) find in M 22 and M 2, respectively, that the 
s--process production may be best fit by pollution from $\ga$3 M$_{\rm \odot}$ 
AGB stars (see also Shingles et al. 2014; Straniero et al. 2014).  In contrast,
Smith et al. (2000) find in $\omega$ Cen that the s--process production may be 
best fit by pollution from $\sim$1.5--3 M$_{\rm \odot}$ AGB stars.  A similar 
conclusion is reached by Marino et al. (2015) through a comparison of the 
change in $\frac{\Delta [Ba/Fe]}{\Delta [Fe/H]}$ between M 22, M 2, and 
NGC 5286.  These authors suggest that the larger [Ba/Fe] range found in M 2 
and NGC 5286 stars compared to M 22 stars may be due to different classes of 
polluters enriching the cluster interstellar mediums.  Although all 
iron--complex clusters exhibit similar r--process dominated metal--poor 
populations, the variable extent of s--process production between clusters may 
be a sign of different enrichment histories.

A key remaining question is whether the s--process--rich populations in 
iron--complex clusters are predominantly influenced by pollution from 
low--intermediate mass AGB stars of the r--process dominated group or 
accretion/mergers from a separate population.  Since most present--day
monometallic globular clusters do not have significant s--process enhancements,
we can speculate that either intracluster pollution or gas accretion from a 
surrounding system (e.g., dwarf galaxy) are the more likely processes.  An
interesting coincidence noted by Marino et al. (2015) is that a large fraction
of the known iron--complex clusters have approximately the same metallicity 
([Fe/H]$\sim$--1.7), and in fact NGC 6273 fits into this pattern.  The 
spatial and kinematic distribution of the iron--complex clusters does not 
strongly suggest that they originated from a common system.  However, 
understanding the significance, if any, of this observation may be an important
component that leads to a better understanding of globular cluster formation.

\subsection{The Metal--rich Anomalous Populations}

As previously noted, the most metal--rich star in NGC 6273 exhibits a 
peculiar composition compared to both the metal--poor and metal--rich 
sub--populations.  Interestingly, the iron--complex clusters $\omega$ Cen, M 2,
and NGC 5286 also host similar peculiar populations.  In all four clusters 
these ``anomalous" stars share the characteristics of being minority 
populations ($\la$5$\%$) that have higher [Fe/H] than the metal--poor 
populations yet have lower [La/Eu] ratios than the metal--rich populations.  
However, the anomalous stars in each cluster are not a universally homogeneous 
population.  For example, the anomalous stars in NGC 6273 and M 2 have low 
[Na/Fe], [Al/Fe], and [$\alpha$/Fe] ratios, those in $\omega$ Cen have an 
unusual O--Na correlation, and those in NGC 5286 have low s--process abundances
but relatively normal light and $\alpha$--element abundances.  The anomalous 
star in NGC 6273 also has low [Cr/Fe] and [Ni/Fe] abundances, which are 
characteristics not shared by any other anomalous population\footnote{We note 
that two of the most metal--rich stars in $\omega$ Cen may have 
[Ni/Fe]$\la$--0.2 (e.g., see Johnson \& Pilachowski 2010, their Figure 10).}. 

The origin of these anomalous metal--rich populations may hold additional clues
for understanding the formation of complex clusters.  
Figures \ref{f11}--\ref{f15} suggest that the anomalous populations of NGC 6273
and M 2 may share a similar composition with the Sagittarius field star 
population.  Therefore, it is tempting to speculate that the minority 
population of anomalous stars present in some clusters is the by product of
accreting field stars from a progenitor system.  Dynamical simulations may
be particularly useful for estimating the fraction of field stars that might
remain bound to the cluster from accretion during such a scenario (e.g., see
Bekki \& Yong 2012).  From an observational stand point, obtaining composition 
information for a larger sample of anomalous stars is needed in order to 
determine if a metallicity spread is present in the anomalous population as 
well, or if those stars represent a monometallic population.  The distinct 
color--magnitude diagram sequences for the anomalous stars in $\omega$ Cen and 
M 2 (e.g., Lee et al. 1999; Milone et al. 2015) suggest that they may be a 
single, coeval population.  However, Marino et al. (2015) note that the 
anomalous stars in NGC 5286 may instead represent a metallicity spread of the 
metal--poor population.  

\begin{figure}
\epsscale{1.00}
\plotone{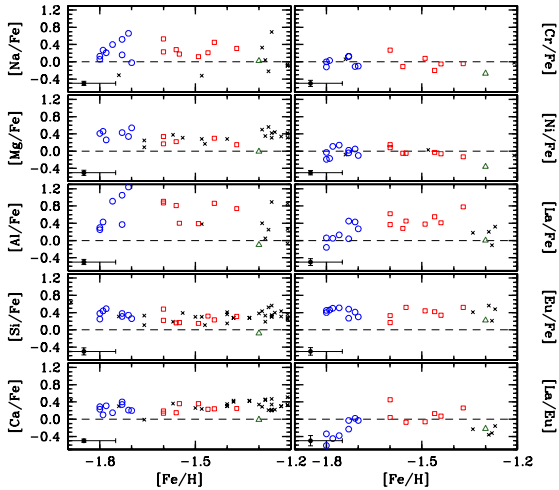}
\caption{Plots of all elements measured for NGC 6273 are shown as a function
of [Fe/H], and the cluster abundances are compared with those of metal--poor
Galactic bulge field stars (black crosses).  The cluster symbols are
the same as those in Figure \ref{f5}.  The bulge field star data are from:
Lecureur et al. (2007), Alves--Brito et al. (2010), McWilliam et al. (2010),
Bensby et al. (2011; 2013), Gonzalez et al. (2011), Johnson et al. (2011; 2012;
2013; 2014), and Garc{\'{\i}}a P{\'e}rez et al. (2013).  Note that the scale
of the ordinate axis is the same for all panels, and that the dashed lines
indicate the solar abundance ratios.}
\label{f16}
\end{figure}

Since NGC 6273 orbits predominantly near the Galactic bulge and between 
$\sim$0.5 and 2.5 kpc from the Galactic center (Carollo et al. 2013; Moreno
et al. 2014), in Figure \ref{f16} we compare the anomalous star's composition 
to the metal--poor bulge field stars.  We find that the anomalous star has 
[Na/Fe], [Al/Fe], [La/Fe], and [Eu/Fe] abundances that are consistent with the 
bulge field star population.  However, the anomalous star's low [$\alpha$/Fe], 
[Cr/Fe], and [Ni/Fe] abundances do not match the bulge field star pattern.
Therefore, we conclude that the anomalous star in NGC 6273 was likely not 
accreted from the present--day bulge field population.

\section{SUMMARY}

Using high resolution spectra obtained with the M2FS instrument on the 
Magellan--Clay telescope, we have measured radial velocities (41 stars) and
chemical compositions (18 stars) for RGB stars in the Galactic globular cluster
NGC 6273.  Our results indicate that NGC 6273 belongs to a growing class of 
``iron--complex" clusters that exhibit both a large [Fe/H] dispersion and a 
correlated increase in [La/Fe] with [Fe/H].  Specifically, we find that 
NGC 6273 hosts at least 2--3 distinct stellar populations: (1) a metal--poor
group with $\langle$[Fe/H]$\rangle$=--1.75 ($\sigma$=0.04) and a spread
in [La/Fe], (2) a metal--rich group with $\langle$[Fe/H]$\rangle$=--1.51 
($\sigma$=0.08) and enhanced [La/Eu] ratios, and (3) a possible anomalous group 
with $\langle$[Fe/H]$\rangle$=--1.30 (1 star) and noticeably lower [X/Fe] 
ratios for nearly all elements.  Despite a moderate sample size, we find that
the two main populations are nearly equivalent in number, and that the 
metal--poor, metal--rich, and anomalous groups constitute 50$\%$, 44$\%$, 
and 6$\%$ of our sample, respectively.  Additionally, the cluster's broad
giant branch may be explained by a combination of significant differential
reddening and a spread in [Fe/H] of at least 0.5 dex.

Interestingly, the metal--poor and metal--rich sub--populations independently
exhibit the well--known Na--Al correlation, but the dispersion in both [Na/Fe] 
and [Al/Fe] is larger for the more metal--poor stars.  Furthermore, the 
[Al/Fe] distribution appears to be largely bimodal, and we do not find any 
stars with $+$0.45$\la$[Al/Fe]$\la$$+$0.75 in either sub--population.  However,
[Na/Fe] and [Al/Fe] are not correlated with either [Fe/H] or [La/Fe], which
suggests that the nucleosynthesis process that creates the Na--Al correlation
operates independently from the process that produces the Fe--peak and 
s--process elements.  The metal--poor and metal--rich groups also exhibit 
similarly enhanced [$\alpha$/Fe] and [Eu/Fe] abundances.

The most unusual composition characteristic of NGC 6273 is the cluster's 
correlated increase in [La/Fe] with [Fe/H].  We find that [La/Fe] increases 
steadily from [La/Fe]$\sim$--0.10 to [La/Fe]$\ga$$+$0.50 between the 
metal--poor and metal--rich groups.  A plot of [La/Eu] versus [La/H] reveals 
that the increase in La abundance is due to almost pure s--process production, 
with perhaps a 5--10$\%$ r--process contribution in the metal--rich group.
Interestingly, the anomalous star has [La/H] and [La/Eu] abundances that are
similar to stars in the metal--poor group.

A comparison of the abundance patterns seen in NGC 6273 with other 
iron--complex clusters (e.g., $\omega$ Cen, M 2, M 22, and NGC 5286) reveals 
that many of them share similar light and heavy element composition 
characteristics.  Furthermore, nearly all iron--complex clusters also have 
[Fe/H]$\sim$--1.7.  In addition to NGC 6273, a few clusters such as $\omega$ 
Cen, M 2, and NGC 5286 also host anomalous metal--rich populations with 
peculiar abundances.  However, the detailed composition for each anomalous 
population may be unique to each cluster, and may indicate that these stars 
were originally part of a different system or accreted from a larger progenitor
host.

Despite growing evidence that NGC 6273 and other iron--complex clusters are 
fundamentally different from monometallic clusters, several key questions 
remain.  One of the most fundamental questions is which physical processes
are required to produce a cluster with and without a metallicity spread?  
Furthermore, why do some iron--complex clusters have a unimodal but broadened
[Fe/H] distribution and others have discreet populations?  Additionally,
why do most (or all) metal--poor clusters with [Fe/H] dispersions have strong
s--process enrichment?  The large [La/Eu] abundances found in iron--complex
clusters are not representative of a typical metal--poor field star nor a
monometallic globular cluster composition.  The nature of the peculiar, 
metal--rich populations seen only in some iron--complex clusters may hold 
additional insight regarding cluster formation.  Specifically, did these
populations form \emph{in situ} with the other metallicity groups, do they
reflect primordial variations, or are they the result of accretion and/or 
merger events from a separate population?  Finally, can more than one formation
and/or evolution process produce clusters with similar composition 
characteristics?

\acknowledgements

This research has made use of NASA's Astrophysics Data System Bibliographic 
Services.  This publication has made use of data products from the Two Micron 
All Sky Survey, which is a joint project of the University of Massachusetts 
and the Infrared Processing and Analysis Center/California Institute of 
Technology, funded by the National Aeronautics and Space Administration and 
the National Science Foundation.  C.I.J. gratefully acknowledges support from 
the Clay Fellowship, administered by the Smithsonian Astrophysical 
Observatory.  R.M.R. acknowledges support from the National Science Foundation 
(AST--1413755 and AST--1412673).  C.A.P. gratefully acknowledges support from 
the Daniel Kirkwood Research Fund at Indiana University and from the National
Science Foundation (AST--1412673).  M.M. is grateful for support from the 
National Science Foundation to develop M2FS (AST--0923160) and carry 
out the observations reported here (AST--1312997), and to the University of 
Michigan for its direct support of M2FS construction and operation.

\clearpage
\setlength{\hoffset}{-0.75in}
\tablenum{1}
\tablecolumns{12}
\tablewidth{0pt}

\begin{deluxetable}{cccccccccccc}
\tabletypesize{\scriptsize}
\tablecaption{Stellar Parameters}
\tablehead{
\colhead{Star Name}	&
\colhead{RA (J2000)}      &
\colhead{DEC (J2000)}      &
\colhead{J}      &
\colhead{H}      &
\colhead{K$_{\rm S}$}      &
\colhead{T$_{\rm eff}$}      &
\colhead{log(g)}      &
\colhead{[Fe/H]}      &
\colhead{$\xi$$_{\rm mic.}$}      &
\colhead{RV$_{\rm helio.}$}      &
\colhead{RV Error}      \\
\colhead{(2MASS)}      &
\colhead{(Degrees)}      &
\colhead{(Degrees)}      &
\colhead{(mag.)}      &
\colhead{(mag.)}      &
\colhead{(mag.)}      &
\colhead{(K)}      &
\colhead{(cgs)}      &
\colhead{(dex)}      &
\colhead{(km s$^{\rm -1}$)}      &
\colhead{(km s$^{\rm -1}$)}      &
\colhead{(km s$^{\rm -1}$)}
}

\startdata
\hline
\multicolumn{12}{c}{Metal--Poor Population}       \\
\hline
17023158$-$2617259	&	255.631607	&	$-$26.290541	&	11.412	&	10.726	&	10.550	&	4400	&	0.80	&	$-$1.80	&	1.85	&	158.72	&	0.24	\\
17023856$-$2617209	&	255.660695	&	$-$26.289145	&	10.980	&	10.287	&	10.134	&	4275	&	0.60	&	$-$1.80	&	1.55	&	143.95	&	0.16	\\
17023293$-$2616127	&	255.637211	&	$-$26.270214	&	11.801	&	11.201	&	10.988	&	4500	&	1.25	&	$-$1.79	&	1.90	&	131.93	&	0.36	\\
17022785$-$2615555	&	255.616065	&	$-$26.265430	&	11.692	&	10.941	&	10.792	&	4425	&	0.90	&	$-$1.78	&	1.80	&	144.57	&	0.19	\\
17024618$-$2615261	&	255.692427	&	$-$26.257250	&	11.284	&	10.518	&	10.322	&	4275	&	0.65	&	$-$1.76	&	1.70	&	162.46	&	0.08	\\
17023289$-$2615535	&	255.637042	&	$-$26.264864	&	11.103	&	10.396	&	10.181	&	4300	&	0.45	&	$-$1.73	&	1.60	&	158.80	&	0.23	\\
17023509$-$2616406	&	255.646228	&	$-$26.277952	&	11.098	&	10.385	&	10.182	&	4225	&	0.65	&	$-$1.73	&	1.70	&	145.77	&	0.19	\\
17023868$-$2616516	&	255.661183	&	$-$26.281012	&	11.323	&	10.572	&	10.400	&	4500	&	1.15	&	$-$1.71	&	1.75	&	138.02	&	0.24	\\
17023384$-$2616416	&	255.641002	&	$-$26.278240	&	11.785	&	11.066	&	10.952	&	4525	&	1.15	&	$-$1.70	&	1.60	&	143.52	&	0.14	\\
\hline
\multicolumn{12}{c}{Metal--Rich Population}       \\
\hline
17024016$-$2616096	&	255.667344	&	$-$26.269346	&	11.416	&	10.699	&	10.451	&	4575	&	1.55	&	$-$1.60	&	1.65	&	126.39	&	0.12	\\
17023078$-$2615183	&	255.628290	&	$-$26.255096	&	11.631	&	10.904	&	10.737	&	4600	&	1.35	&	$-$1.60	&	1.90	&	157.43	&	0.06	\\
17025121$-$2617230	&	255.713406	&	$-$26.289745	&	11.079	&	10.280	&	10.104	&	4450	&	1.15	&	$-$1.56	&	1.75	&	134.13	&	0.37	\\
17024326$-$2617504	&	255.680281	&	$-$26.297361	&	11.864	&	11.166	&	10.982	&	4450	&	1.20	&	$-$1.55	&	1.60	&	123.37	&	0.22	\\
17025221$-$2614307	&	255.717545	&	$-$26.241865	&	11.807	&	11.118	&	10.972	&	4400	&	1.20	&	$-$1.49	&	1.45	&	148.81	&	0.30	\\
17023424$-$2615437	&	255.642703	&	$-$26.262144	&	10.977	&	10.215	&	10.000	&	4325	&	1.25	&	$-$1.46	&	1.75	&	153.76	&	0.07	\\
17023481$-$2617152	&	255.645044	&	$-$26.287563	&	11.415	&	10.681	&	10.535	&	4450	&	1.20	&	$-$1.44	&	1.80	&	148.11	&	0.20	\\
17023301$-$2615360	&	255.637556	&	$-$26.260017	&	11.211	&	10.482	&	10.264	&	4400	&	1.65	&	$-$1.37	&	1.90	&	141.00	&	0.25	\\
\hline
\multicolumn{12}{c}{Anomalous Population}       \\
\hline
17024453$-$2616377	&	255.685573	&	$-$26.277155	&	11.288	&	10.507	&	10.307	&	4350	&	1.10	&	$-$1.30	&	1.60	&	143.06	&	0.21	\\
\enddata

\end{deluxetable}

\clearpage
\setlength{\voffset}{-1.70in}
\tablenum{2}
\tablecolumns{5}
\tablewidth{0pt}

\begin{deluxetable}{ccccc}
\tablecaption{Line list and Adopted Solar Abundances}
\tabletypesize{\footnotesize}
\tablehead{
\colhead{Wavelength}	&
\colhead{Species}      &
\colhead{E.P.}      &
\colhead{log(gf)\tablenotemark{a}}      &
\colhead{log $\epsilon$(X)$_{\rm \odot}$}      \\
\colhead{(\AA)}	&
\colhead{}      &
\colhead{(eV)}      &
\colhead{}      &
\colhead{}      
}

\startdata
6154.23 &       Na I    &       2.10    &       $-$1.560        &       6.31    \\
6160.75 &       Na I    &       2.10    &       $-$1.210        &       6.31    \\
6318.71 &       Mg I    &       5.10    &       $-$2.010        &       7.58    \\
6319.24 &       Mg I    &       5.10    &       $-$2.250        &       7.58    \\
6319.49 &       Mg I    &       5.10    &       $-$2.730        &       7.58    \\
6696.02 &       Al I    &       3.14    &       $-$1.520        &       6.45    \\
6698.67 &       Al I    &       3.14    &       $-$1.910        &       6.45    \\
6142.48 &       Si I    &       5.62    &       $-$1.565        &       7.55    \\
6155.13 &       Si I    &       5.62    &       $-$0.764        &       7.55    \\
6237.32 &       Si I    &       5.61    &       $-$0.965        &       7.55    \\
6407.29 &       Si I    &       5.87    &       $-$1.353        &       7.55    \\
6414.98 &       Si I    &       5.87    &       $-$1.055        &       7.55    \\
6166.44 &       Ca I    &       2.52    &       $-$1.182        &       6.36    \\
6169.04 &       Ca I    &       2.52    &       $-$0.717        &       6.36    \\
6169.56 &       Ca I    &       2.53    &       $-$0.538        &       6.36    \\
6455.6  &       Ca I    &       2.52    &       $-$1.374        &       6.36    \\
6471.66 &       Ca I    &       2.53    &       $-$0.726        &       6.36    \\
6499.65 &       Ca I    &       2.52    &       $-$0.858        &       6.36    \\
6330.09 &       Cr I    &       0.94    &       $-$3.000        &       5.67    \\
6501.19 &       Cr I    &       0.98    &       $-$3.965        &       5.67    \\
6630.01 &       Cr I    &       1.03    &       $-$3.570        &       5.67    \\
6151.62 &       Fe I    &       2.18    &       $-$3.379        &       7.52    \\
6159.37 &       Fe I    &       4.61    &       $-$1.950        &       7.52    \\
6165.36 &       Fe I    &       4.14    &       $-$1.584        &       7.52    \\
6173.33 &       Fe I    &       2.22    &       $-$2.930        &       7.52    \\
6180.20 &       Fe I    &       2.73    &       $-$2.629        &       7.52    \\
6187.99 &       Fe I    &       3.94    &       $-$1.690        &       7.52    \\
6191.56 &       Fe I    &       2.43    &       $-$1.727        &       7.52    \\
6199.51 &       Fe I    &       2.56    &       $-$4.360        &       7.52    \\
6200.31 &       Fe I    &       2.61    &       $-$2.437        &       7.52    \\
6213.43 &       Fe I    &       2.22    &       $-$2.692        &       7.52    \\
6219.28 &       Fe I    &       2.20    &       $-$2.563        &       7.52    \\
6220.78 &       Fe I    &       3.88    &       $-$2.420        &       7.52    \\
6229.23 &       Fe I    &       2.85    &       $-$2.885        &       7.52    \\
6230.72 &       Fe I    &       2.56    &       $-$1.291        &       7.52    \\
6232.64 &       Fe I    &       3.65    &       $-$1.263        &       7.52    \\
6240.65 &       Fe I    &       2.22    &       $-$3.353        &       7.52    \\
6246.32 &       Fe I    &       3.60    &       $-$0.773        &       7.52    \\
6252.56 &       Fe I    &       2.40    &       $-$1.847        &       7.52    \\
6253.83 &       Fe I    &       4.73    &       $-$1.500        &       7.52    \\
6270.22 &       Fe I    &       2.86    &       $-$2.649        &       7.52    \\
6271.28 &       Fe I    &       3.33    &       $-$2.783        &       7.52    \\
6315.81 &       Fe I    &       4.08    &       $-$1.720        &       7.52    \\
6322.69 &       Fe I    &       2.59    &       $-$2.446        &       7.52    \\
6330.85 &       Fe I    &       4.73    &       $-$1.230        &       7.52    \\
6335.33 &       Fe I    &       2.20    &       $-$2.387        &       7.52    \\
6336.82 &       Fe I    &       3.69    &       $-$0.866        &       7.52    \\
6380.74 &       Fe I    &       4.19    &       $-$1.376        &       7.52    \\
6385.72 &       Fe I    &       4.73    &       $-$1.840        &       7.52    \\
6392.54 &       Fe I    &       2.28    &       $-$4.010        &       7.52    \\
6393.60 &       Fe I    &       2.43    &       $-$1.676        &       7.52    \\
6411.65 &       Fe I    &       3.65    &       $-$0.625        &       7.52    \\
6419.95 &       Fe I    &       4.73    &       $-$0.340        &       7.52    \\
6421.35 &       Fe I    &       2.28    &       $-$2.017        &       7.52    \\
6430.85 &       Fe I    &       2.18    &       $-$2.066        &       7.52    \\
6436.41 &       Fe I    &       4.19    &       $-$2.340        &       7.52    \\
6469.19 &       Fe I    &       4.83    &       $-$0.690        &       7.52    \\
6481.87 &       Fe I    &       2.28    &       $-$2.814        &       7.52    \\
6496.47 &       Fe I    &       4.79    &       $-$0.680        &       7.52    \\
6498.94 &       Fe I    &       0.96    &       $-$4.629        &       7.52    \\
6518.37 &       Fe I    &       2.83    &       $-$2.550        &       7.52    \\
6591.31 &       Fe I    &       4.59    &       $-$2.130        &       7.52    \\
6592.91 &       Fe I    &       2.73    &       $-$1.723        &       7.52    \\
6593.87 &       Fe I    &       2.43    &       $-$2.462        &       7.52    \\
6597.56 &       Fe I    &       4.79    &       $-$1.000        &       7.52    \\
6609.11 &       Fe I    &       2.56    &       $-$2.632        &       7.52    \\
6627.54 &       Fe I    &       4.55    &       $-$1.530        &       7.52    \\
6633.41 &       Fe I    &       4.83    &       $-$1.240        &       7.52    \\
6633.75 &       Fe I    &       4.56    &       $-$0.849        &       7.52    \\
6646.93 &       Fe I    &       2.61    &       $-$4.000        &       7.52    \\
6703.57 &       Fe I    &       2.76    &       $-$3.080        &       7.52    \\
6705.10 &       Fe I    &       4.61    &       $-$1.102        &       7.52    \\
6149.26 &       Fe II   &       3.89    &       $-$2.681        &       7.52    \\
6247.56 &       Fe II   &       3.89    &       $-$2.245        &       7.52    \\
6416.92 &       Fe II   &       3.89    &       $-$2.627        &       7.52    \\
6432.68 &       Fe II   &       2.89    &       $-$3.547        &       7.52    \\
6456.38 &       Fe II   &       3.90    &       $-$2.115        &       7.52    \\
6516.08 &       Fe II   &       2.89    &       $-$3.142        &       7.52    \\
6175.36 &       Ni I    &       4.09    &       $-$0.469        &       6.25    \\
6176.81 &       Ni I    &       4.09    &       $-$0.240        &       6.25    \\
6177.24 &       Ni I    &       1.83    &       $-$3.460        &       6.25    \\
6191.17 &       Ni I    &       1.68    &       $-$2.303        &       6.25    \\
6223.98 &       Ni I    &       4.11    &       $-$0.820        &       6.25    \\
6378.25 &       Ni I    &       4.15    &       $-$0.760        &       6.25    \\
6598.59 &       Ni I    &       4.24    &       $-$0.970        &       6.25    \\
6262.29 &       La II   &       0.40    &       hfs     &       1.13    \\
6390.48 &       La II   &       0.32    &       hfs     &       1.13    \\
6437.64 &       Eu II   &       1.32    &       hfs     &       0.52    \\
6645.06 &       Eu II   &       1.38    &       hfs     &       0.52    \\
\enddata

\tablenotetext{a}{The ``hfs" designation indicates the abundance was
calculated taking hyperfine structure into account.  See text for details.}

\end{deluxetable}

\clearpage
\tablenum{3a}
\tablecolumns{11}
\tablewidth{0pt}

\begin{deluxetable}{ccccccccccc}
\tablecaption{Stellar Abundances and Uncertainties}
\tabletypesize{\footnotesize}
\tablehead{
\colhead{Star Name}	&
\colhead{[Fe/H]}      &
\colhead{$\Delta$[Fe/H]}      &
\colhead{[Na/Fe]}      &
\colhead{$\Delta$[Na/Fe]}      &
\colhead{[Mg/Fe]}      &
\colhead{$\Delta$[Mg/Fe]}      &
\colhead{[Al/Fe]}      &
\colhead{$\Delta$[Al/Fe]}      &
\colhead{[Si/Fe]}      &
\colhead{$\Delta$[Si/Fe]}       
}

\startdata
\hline
\multicolumn{11}{c}{Metal--Poor Population}       \\
\hline
17023158$-$2617259	&	$-$1.80	&	0.10	&	$+$0.13	&	0.05	&	$+$0.41	&	0.06	&	$+$0.30	&	0.05	&	$+$0.38	&	0.07	\\
17023856$-$2617209	&	$-$1.80	&	0.11	&	$+$0.06	&	0.05	&	\nodata	&	\nodata	&	$+$0.25	&	0.05	&	$+$0.25	&	0.09	\\
17023293$-$2616127	&	$-$1.79	&	0.09	&	$+$0.27	&	0.05	&	$+$0.46	&	0.06	&	$+$0.43	&	0.05	&	$+$0.45	&	0.08	\\
17022785$-$2615555	&	$-$1.78	&	0.10	&	$+$0.21	&	0.05	&	$+$0.26	&	0.06	&	\nodata	&	\nodata	&	$+$0.49	&	0.08	\\
17024618$-$2615261	&	$-$1.76	&	0.11	&	$+$0.40	&	0.07	&	\nodata	&	\nodata	&	$+$0.91	&	0.05	&	\nodata	&	\nodata	\\
17023289$-$2615535	&	$-$1.73	&	0.11	&	$+$0.52	&	0.04	&	$+$0.43	&	0.06	&	$+$1.05	&	0.06	&	$+$0.31	&	0.08	\\
17023509$-$2616406	&	$-$1.73	&	0.11	&	$+$0.16	&	0.05	&	\nodata	&	\nodata	&	$+$0.37	&	0.05	&	$+$0.38	&	0.08	\\
17023868$-$2616516	&	$-$1.71	&	0.10	&	$+$0.66	&	0.03	&	$+$0.35	&	0.06	&	$+$1.24	&	0.03	&	$+$0.34	&	0.08	\\
17023384$-$2616416	&	$-$1.70	&	0.10	&	$-$0.02	&	0.05	&	$+$0.54	&	0.06	&	\nodata	&	\nodata	&	$+$0.26	&	0.06	\\
\hline
\multicolumn{11}{c}{Metal--Rich Population}       \\
\hline
17024016$-$2616096	&	$-$1.60	&	0.10	&	$+$0.23	&	0.05	&	$+$0.34	&	0.05	&	$+$0.88	&	0.03	&	$+$0.22	&	0.11	\\
17023078$-$2615183	&	$-$1.60	&	0.09	&	$+$0.53	&	0.05	&	$+$0.17	&	0.05	&	$+$0.91	&	0.07	&	$+$0.48	&	0.06	\\
17025121$-$2617230	&	$-$1.56	&	0.10	&	$+$0.28	&	0.07	&	$+$0.22	&	0.06	&	$+$0.81	&	0.05	&	$+$0.16	&	0.08	\\
17024326$-$2617504	&	$-$1.55	&	0.10	&	$+$0.18	&	0.06	&	\nodata	&	\nodata	&	$+$0.40	&	0.05	&	$+$0.17	&	0.11	\\
17025221$-$2614307	&	$-$1.49	&	0.11	&	$+$0.12	&	0.05	&	\nodata	&	\nodata	&	$+$0.39	&	0.04	&	$+$0.15	&	0.10	\\
17023424$-$2615437	&	$-$1.46	&	0.10	&	$+$0.21	&	0.05	&	\nodata	&	\nodata	&	\nodata	&	\nodata	&	$+$0.32	&	0.08	\\
17023481$-$2617152	&	$-$1.44	&	0.11	&	$+$0.46	&	0.04	&	$+$0.30	&	0.06	&	$+$0.87	&	0.09	&	$+$0.23	&	0.08	\\
17023301$-$2615360	&	$-$1.37	&	0.12	&	$+$0.31	&	0.05	&	$+$0.15	&	0.05	&	$+$0.74	&	0.05	&	$+$0.31	&	0.08	\\
\hline
\multicolumn{11}{c}{Anomalous Population}       \\
\hline
17024453$-$2616377	&	$-$1.30	&	0.11	&	$+$0.02	&	0.09	&	$-$0.01	&	0.06	&	$-$0.10	&	0.11	&	$-$0.08	&	0.09	\\
\enddata

\end{deluxetable}

\clearpage
\tablenum{3b}
\tablecolumns{11}
\tablewidth{0pt}

\begin{deluxetable}{ccccccccccc}
\tabletypesize{\footnotesize}
\tablecaption{Stellar Abundances and Uncertainties}
\tablehead{
\colhead{Star Name}	&
\colhead{[Ca/Fe]}      &
\colhead{$\Delta$[Ca/Fe]}      &
\colhead{[Cr/Fe]}      &
\colhead{$\Delta$[Cr/Fe]}      &
\colhead{[Ni/Fe]}      &
\colhead{$\Delta$[Ni/Fe]}      &
\colhead{[La/Fe]}      &
\colhead{$\Delta$[La/Fe]}      &
\colhead{[Eu/Fe]}      &
\colhead{$\Delta$[Eu/Fe]}      
}

\startdata
\hline
\multicolumn{11}{c}{Metal--Poor Population}       \\
\hline
17023158$-$2617259	&	$+$0.23	&	0.02	&	$+$0.00	&	0.13	&	$-$0.19	&	0.06	&	$-$0.16	&	0.08	&	$+$0.45	&	0.10	\\
17023856$-$2617209	&	$+$0.29	&	0.03	&	$-$0.12	&	0.06	&	$-$0.03	&	0.05	&	$+$0.06	&	0.08	&	$+$0.40	&	0.10	\\
17023293$-$2616127	&	$+$0.10	&	0.05	&	$+$0.03	&	0.05	&	$-$0.17	&	0.06	&	\nodata	&	\nodata	&	$+$0.47	&	0.09	\\
17022785$-$2615555	&	$+$0.31	&	0.02	&	\nodata	&	\nodata	&	$+$0.11	&	0.06	&	$+$0.05	&	0.08	&	$+$0.50	&	0.09	\\
17024618$-$2615261	&	$+$0.15	&	0.06	&	\nodata	&	\nodata	&	$+$0.14	&	0.03	&	$+$0.13	&	0.08	&	$+$0.51	&	0.09	\\
17023289$-$2615535	&	$+$0.34	&	0.05	&	$+$0.13	&	0.06	&	$-$0.04	&	0.05	&	$+$0.04	&	0.09	&	$+$0.28	&	0.10	\\
17023509$-$2616406	&	$+$0.40	&	0.05	&	$+$0.12	&	0.12	&	$+$0.02	&	0.04	&	$+$0.45	&	0.08	&	$+$0.48	&	0.09	\\
17023868$-$2616516	&	$+$0.21	&	0.06	&	$-$0.11	&	0.05	&	$+$0.05	&	0.03	&	$+$0.43	&	0.07	&	$+$0.41	&	0.09	\\
17023384$-$2616416	&	$+$0.20	&	0.03	&	$-$0.10	&	0.05	&	$-$0.10	&	0.02	&	$+$0.27	&	0.08	&	$+$0.30	&	0.09	\\
\hline
\multicolumn{11}{c}{Metal--Rich Population}       \\
\hline
17024016$-$2616096	&	$+$0.19	&	0.04	&	\nodata	&	\nodata	&	$+$0.08	&	0.05	&	$+$0.37	&	0.07	&	$+$0.33	&	0.09	\\
17023078$-$2615183	&	$+$0.14	&	0.05	&	$+$0.27	&	0.05	&	$+$0.15	&	0.07	&	$+$0.62	&	0.08	&	$+$0.17	&	0.09	\\
17025121$-$2617230	&	$+$0.15	&	0.04	&	$-$0.11	&	0.06	&	$-$0.05	&	0.06	&	$+$0.28	&	0.08	&	\nodata	&	\nodata	\\
17024326$-$2617504	&	$+$0.36	&	0.03	&	\nodata	&	\nodata	&	$-$0.05	&	0.04	&	$+$0.45	&	0.07	&	$+$0.52	&	0.08	\\
17025221$-$2614307	&	$+$0.36	&	0.04	&	$+$0.08	&	0.06	&	\nodata	&	\nodata	&	$+$0.39	&	0.07	&	$+$0.44	&	0.09	\\
17023424$-$2615437	&	$+$0.23	&	0.03	&	$-$0.20	&	0.08	&	$-$0.03	&	0.05	&	$+$0.55	&	0.07	&	$+$0.42	&	0.07	\\
17023481$-$2617152	&	$+$0.24	&	0.03	&	$-$0.05	&	0.06	&	$-$0.06	&	0.07	&	$+$0.41	&	0.08	&	$+$0.34	&	0.09	\\
17023301$-$2615360	&	$+$0.25	&	0.08	&	$-$0.04	&	0.07	&	$-$0.13	&	0.08	&	$+$0.78	&	0.07	&	$+$0.52	&	0.08	\\
\hline
\multicolumn{11}{c}{Anomalous Population}       \\
\hline
17024453$-$2616377	&	$-$0.01	&	0.04	&	$-$0.27	&	0.07	&	$-$0.36	&	0.05	&	$+$0.00	&	0.08	&	$+$0.22	&	0.09	\\
\enddata

\end{deluxetable}

\clearpage
\tablenum{4}
\tablecolumns{8}
\tablewidth{0pt}

\begin{deluxetable}{cccccccc}
\tablecaption{Heliocentric Radial Velocities for Additional Stars}
\tablehead{
\colhead{Star Name}	&
\colhead{RA (J2000)}      &
\colhead{DEC (J2000)}      &
\colhead{J}      &
\colhead{H}      &
\colhead{K$_{\rm S}$}      &
\colhead{RV$_{\rm helio.}$}      &
\colhead{RV Error}      \\
\colhead{(2MASS)}      &
\colhead{(Degrees)}      &
\colhead{(Degrees)}      &
\colhead{(mag.)}      &
\colhead{(mag.)}      &
\colhead{(mag.)}      &
\colhead{(km s$^{\rm -1}$)}      &
\colhead{(km s$^{\rm -1}$)}
}

\startdata
\hline
\multicolumn{8}{c}{Radial Velocity Members}       \\
\hline
17022040$-$2616289	&	255.585022	&	$-$26.274719	&	11.532	&	10.892	&	10.744	&	148.20	&	0.20	\\
17022395$-$2614538	&	255.599796	&	$-$26.248289	&	11.721	&	11.061	&	10.898	&	148.37	&	0.19	\\
17022878$-$2614320	&	255.619947	&	$-$26.242231	&	11.304	&	10.556	&	10.393	&	145.56	&	0.22	\\
17023286$-$2616475	&	255.636949	&	$-$26.279886	&	11.258	&	10.594	&	10.410	&	142.74	&	0.23	\\
17023592$-$2615595	&	255.649707	&	$-$26.266548	&	11.279	&	10.549	&	10.370	&	162.37	&	0.26	\\
17023649$-$2615229	&	255.652083	&	$-$26.256386	&	11.327	&	10.582	&	10.429	&	156.82	&	0.33	\\
17023685$-$2616217	&	255.653580	&	$-$26.272709	&	11.765	&	11.110	&	10.883	&	130.58	&	0.67	\\
17023811$-$2617392	&	255.658832	&	$-$26.294239	&	11.678	&	11.011	&	10.890	&	138.03	&	0.22	\\
17023874$-$2612434	&	255.661456	&	$-$26.212059	&	11.396	&	10.737	&	10.542	&	145.84	&	0.20	\\
17023946$-$2615017	&	255.664417	&	$-$26.250486	&	11.848	&	11.237	&	10.998	&	136.15	&	1.16	\\
17023984$-$2617360	&	255.666020	&	$-$26.293348	&	11.535	&	10.851	&	10.647	&	157.02	&	0.08	\\
17024104$-$2616507	&	255.671004	&	$-$26.280752	&	11.803	&	11.169	&	10.990	&	142.40	&	0.31	\\
17024132$-$2613517	&	255.672190	&	$-$26.231030	&	11.706	&	11.010	&	10.876	&	137.69	&	0.11	\\
17024165$-$2617033	&	255.673565	&	$-$26.284258	&	11.239	&	10.563	&	10.359	&	150.92	&	0.43	\\
17024168$-$2615100	&	255.673699	&	$-$26.252796	&	11.662	&	10.985	&	10.761	&	132.44	&	1.41	\\
17024289$-$2615274	&	255.678722	&	$-$26.257622	&	11.501	&	10.830	&	10.652	&	138.28	&	0.28	\\
17024371$-$2620183	&	255.682141	&	$-$26.338444	&	11.811	&	11.111	&	10.980	&	148.52	&	0.04	\\
17024416$-$2615177	&	255.684021	&	$-$26.254919	&	11.384	&	10.685	&	10.502	&	136.60	&	0.35	\\
17024566$-$2615124	&	255.690259	&	$-$26.253452	&	11.160	&	10.457	&	10.221	&	137.47	&	0.15	\\
17024717$-$2615107	&	255.696559	&	$-$26.252991	&	11.566	&	10.887	&	10.676	&	151.45	&	0.05	\\
17025033$-$2615582	&	255.709727	&	$-$26.266191	&	11.667	&	10.860	&	10.714	&	143.98	&	0.37	\\
\hline
\multicolumn{8}{c}{Radial Velocity Non--Members}       \\
\hline
17023847$-$2618509\tablenotemark{a}	&	255.660330	&	$-$26.314159	&	11.067	&	10.354	&	10.213	&	$-$31.98	&	0.20	\\
17024093$-$2620182	&	255.670556	&	$-$26.338413	&	11.462	&	10.786	&	10.622	&	15.25	&	0.21	\\
\enddata

\tablenotetext{a}{This star exhibits a double--lined spectrum with one 
component yielding a velocity consistent with cluster membership and the other
component yielding a velocity inconsistent with cluster membership.}

\end{deluxetable}

\clearpage
\setlength{\hoffset}{-0.75in}
\tablenum{5}
\tablecolumns{12}
\tablewidth{0pt}

\begin{deluxetable}{cccccccccccc}
\tablecaption{Intracluster Comparison}
\tabletypesize{\scriptsize}
\tablehead{
\colhead{Statistic}     &
\colhead{[Fe/H]}     &
\colhead{[Na/Fe]}     &
\colhead{[Mg/Fe]}     &
\colhead{[Al/Fe]}     &
\colhead{[Si/Fe]}     &
\colhead{[Ca/Fe]}     &
\colhead{[Cr/Fe]}     &
\colhead{[Ni/Fe]}     &
\colhead{[La/Fe]}     &
\colhead{[Eu/Fe]}     &
\colhead{RV$_{\rm helio.}$}
}

\startdata
\hline
\multicolumn{12}{c}{Metal--Poor Population}       \\
\hline
Average	   &	 $-$1.75	&	  $+$0.26	 &	  $+$0.41	 &	  $+$0.65	 &	  $+$0.35	 &	  $+$0.25	 &	 $-$0.01	 &	 $-$0.02	 &	 $+$0.16	 &	 $+$0.42	 &	  $+$147.53	 \\
$\sigma$	   &	0.04	&	0.22	 &	0.10	 &	0.40	 &	0.09	 &	0.10	 &	0.10	 &	0.12	 &	0.21	 &	0.08	 &	10.30	 \\
\hline
\multicolumn{12}{c}{Metal--Rich Population}       \\
\hline
Average	       &	 $-$1.51	&	  $+$0.29	 &	  $+$0.24	 &	  $+$0.71	 &	  $+$0.25	 &	  $+$0.24	 &	 $-$0.01	 &	 $-$0.01	 &	 $+$0.48	 &	 $+$0.39	 &	   $+$141.63	 \\
$\sigma$	       &	0.08	&	0.14	 &	0.08	 &	0.22	 &	0.11	 &	0.08	 &	0.16	 &	0.09	 &	0.16	 &	0.12	 &	12.61	 \\
\hline
\multicolumn{12}{c}{Anomalous Population}       \\
\hline
Average	   &	$-$1.30	&	$+$0.02	&	$-$0.01	&	$-$0.10	&	$-$0.08	&	$-$0.01	&	$-$0.27	&	$-$0.36	&	$+$0.00	&	$+$0.22	&	$+$143.06	\\
$\sigma$	   &	\nodata	&	\nodata	&	\nodata	&	\nodata	&	\nodata	&	\nodata	&	\nodata	&	\nodata	&	\nodata	&	\nodata	&	\nodata	\\
\enddata

\end{deluxetable}


\end{document}